\documentclass[journal,12pt,onecolumn,draftclsnofoot,]{IEEEtranTCOM}
\normalsize
\usepackage[pdftex]{graphicx}
  \usepackage{epstopdf}
  \DeclareGraphicsExtensions{.pdf,.jpeg,.png,.eps}
\usepackage{color}
\usepackage[cmex10]{amsmath}
\usepackage{amssymb}
\usepackage{algorithmic}
\usepackage{array}
\usepackage{mdwmath}
\usepackage{amsthm}
\usepackage{mdwtab}
\usepackage{tabularx}
\usepackage{stfloats}
\usepackage{cite}
\makeatother

\allowdisplaybreaks
\newtheoremstyle{mystyle}
  {3pt}
  {3pt}
  {\itshape} 
  {\parindent}
  {\bfseries}
  {\upshape{:}}
  {.5em}
  {}

\theoremstyle{mystyle}  
\newtheorem{theorem}{Theorem}

\theoremstyle{mystyle}  
\newtheorem{coro}{Corollary}
\theoremstyle{mystyle}
\newtheorem{prop}{Proposition}
\usepackage{subcaption}
\usepackage[font={small}]{caption}

\begin{document}
\title{Unif\mbox{}ied Analysis of SWIPT Relay Networks with Noncoherent Modulation}
\author{Lina Mohjazi,~\IEEEmembership{Student Member,~IEEE,} Sami Muhaidat,~\IEEEmembership{Senior Member,~IEEE,} Mehrdad Dianati,~\IEEEmembership{Senior Member,~IEEE,} and Mahmoud Al-Qutayri,~\IEEEmembership{Senior Member,~IEEE}
\thanks {This work was accepted in part at the IEEE VTC Fall, Toronto, Canada, 2017 \cite{Mohjazi3}.}
 \thanks{L. Mohjazi and S. Muhaidat are with the Institute for Communication Systems, Department of Electronic Engineering, University of Surrey, Surrey, U.K. (e-mail: l.mohjazi@surrey.ac.uk, muhaidat@ieee.org).} 
\thanks{ M. Dianati is with the Warwick Manufacturing Group, University of Warwick, Coventry, U.K. (e-mail: m.dianati@warwick.ac.uk).}
\thanks{S. Muhaidat and M. Al-Qutayri are with the Department of Electrical and Computer Engineering, Khalifa University, Abu Dhabi, UAE (e-mail: muhaidat@ieee.org, mqutayri@kustar.ac.ae).}}

\maketitle

\vspace*{-2cm}
\markboth{Submitted to IEEE TRANSACTIONS on WIRELESS COMMUNICATIONS}{}

\begin{abstract}
Simultaneous wireless information and power transfer (SWIPT) relay networks represent a promising in the development of wireless networks, enabling simultaneous radio frequency (RF) energy harvesting (EH) and information processing. Different from conventional SWIPT relaying schemes, which use coherent modulation that typically assume the availability of perfect channel state information (CSI), in this work, we consider the application of \textit{noncoherent} modulation in order to avoid the need of instantaneous CSI estimation/tracking and minimise energy consumption.  We propose a unif\mbox{}ied and comprehensive analytical framework for the analysis of time switching (TS) and power splitting (PS) receiver architectures with the amplify-and-forward (AF) relaying protocol. In particular, we adopt a moments-based approach to derive novel expressions for the outage probability, achievable throughput, and average symbol error rate (ASER) of a dual-hop SWIPT relay system. Furthermore, we derive new asymptotic analytical results for the outage probability and ASER in the high SNR regime and we analytically quantify the achievable diversity order of the considered system. We analyse the impact of several system parameters, involving the energy conversion efficiency and TS and PS ratio assumptions, imposed on the EH relay terminal. An extensive Monte Carlo simulation study is presented to corroborate the proposed analytical model.
\end{abstract}

\begin{IEEEkeywords}
\vspace*{-0.5cm}
Wireless information and power transfer, energy harvesting, \textit{noncoherent} modulation, relaying networks, symbol-error-rate (SER), outage probability, performance analysis.
\end{IEEEkeywords}
\vspace*{-0.5cm}
\section{Introduction}
\IEEEPARstart{S}{IMULTANEOUS} wireless information and power transfer (SWIPT) is an emerging technology which was proposed to prolong the lifetime of energy-constrained wireless networks and to provide an alternative solution to limited lifetime battery-operated devices. In this context, designing efficient transmission mechanisms and protocols for SWIPT has been the recent focus of intensive research, due to its promising applications in the next generation  wireless networks \cite{Bi,Mohjazi2,Lmohjazi}. With  SWIPT, a wireless terminal is powered up by a received Radio Frequency (RF) signal and, simultaneously, information processing is carried out using the same signal \cite{Varshney2008,Grover}. 
\par SWIPT-based relaying was proposed as a promising technique to provide throughput improvement, communication reliability enhancement, and coverage range extension \cite{Nasir2013}. From this perspective, the theoretical and implementation aspects of SWIPT relay networks have been areas of active research interest \cite{Rabie,Zhiguo1,Krikidis,Lee}. A SWIPT-enabled relay terminal harvests energy from the signal transmitted from the source node to forward the information to a destination node. Owing to practical hardware limitations, a relay node is unable to perform the two functions of energy harvesting (EH) and information processing simultaneously. To address this problem, two practically realisable SWIPT relaying protocols were proposed in \cite{Nasir2013}, namely, time switching (TS) and power splitting (PS) relaying protocols. In the former, the relay switches over time between EH and information processing, whereas, in the latter, the relay uses portion of the received power for EH and the remaining for information processing. Due to the unique characteristics inherent in such energy-constrained relay nodes, minimising the total energy consumption is crucial for the design of SWIPT relay systems. 
\par Towards achieving that, an energy-efficient modulation scheme with a low-complexity implementation that is still robust enough to provide the desired service has to be appropriately devised. Although there has been a growing literature on SWIPT, particularly, in the context of relay networks, see e.g., \cite{Rabie,Zhiguo1,Krikidis,Lee} and the references therein, all research studies were built upon the assumption of perfect channel state information (CSI) knowledge to allow for \textit{coherent} information delivery. However, in such scenarios, the source is required to periodically send training symbols which are then forwarded to the destination through the relay node. Additionally, in some cases, the relay may need to estimate the source-relay channels \cite{Liu2010}. 
Although CSI-based relaying systems are expected to outperform those that depend on the absence of CSI, this improvement comes at the inevitable cost of increased signaling overhead and processing burden. This, in turn, increases the amount of power consumption at the relay node and, hence, poses a practical hindrance in the performance of SWIPT relay networks. Thus, eliminating the need of CSI estimation appears to be an interesting proposition in this context.  
\par Recently, \textit{noncoherent} modulation techniques for SWIPT relay systems have been proposed in \cite{Liu, Gazor,Xu,Mohjazi,Lou} to circumvent channel estimation. In particular, in \cite{Liu, Gazor}, maximum-likelihood detectors were obtained based on TS and PS SWIPT receiver architectures, for amplify-and-forward (AF) and decode-and-forward (DF) relaying, respectively. The performance of the proposed receivers in terms of their average symbol error rate (ASER) was studied via Monte-Carlo simulations only. Furthermore, \textit{noncoherent} modulation for two-way relay networks was investigated in \cite{Xu}. In \cite{Mohjazi}, we derived bounds on the average bit error rate (ABER) of SWIPT AF relay networks, employing the PS relaying protocol, for binary differential phase-shift keying (BDPSK). In our analysis, we ignored the dependency of the relay-to-destination channel on the source-to-relay channel resulting from the PS at the relay. Our results demonstrated that the bound is tight in the low to medium signal-to-noise (SNR) regime and that there is no loss in the ABER performance in this region. In \cite{Lou}, exact closed-form ABER expression of a selection combining (SC) scheme was derived for a differential cooperative system employing an AF relay with SWIPT capability.
\par \textit{Although analysing the performance and understanding the limitations of \textit{noncoherent} SWIPT relay networks in various operational conditions is essential for their design, implementation, testing and deployment stages, to the best of our knowledge, none of the previous works provided a comprehensive and unif\mbox{}ied analytical treatment to efficiently evaluate the performance of such systems}. This stems from the fact that finding exact closed-form expressions for the probability density function (PDF) and cumulative distribution function (CDF) of the receive SNR of SWIPT relay networks is a challenging task.   
\par Motivated by this, the aim of this paper is to fill this research gap by developing a comprehensive and unif\mbox{}ied mathematical framework to analytically study the performance of arbitrary $M$-ary \textit{noncoherent} frequency-shift keying ($M$-FSK) and $M$-ary DPSK ($M$-DPSK) in dual-hop AF relaying protocols with TS and PS SWIPT schemes. Different from the work presented in \cite{Mohjazi}, in this work, we consider the assumption of a general SWIPT relay system, where the channels of the two hops are correlated. We address the challenge of deriving the PDF of the receive SNR by alternatively resorting to the moments-based approach. 
\par Specifically, the main contributions and results of this paper are summarised as follows:
\begin{itemize}
\item We derive a computationally effective exact closed-form expression for the moments of the receive SNR when the channels of the two hops are subject to Rayleigh fading. To the best of our knowledge, this result is novel in literature.
\item We employ the Pad$\acute{\text{e}}$ approximation (PA) technique to obtain an accurate approximate rational expression of the corresponding moment generating function (MGF) of the system based on the knowledge of its moments.
\item The derived moments and MGF expressions are utilised to evaluate fundamental performance metrics such as, the average SNR (ASNR), the amount-of-fading (AoF), the outage probability, the achievable throughput, and the ASER for the cases of arbitrary $M$-FSK and $M$-DPSK.
\item We derive new asymptotic expressions for the outage probability and the ASER, respectively.
\item Through the derived outage probability expression, we demonstrate that the achievable diversity order of the considered system model is less than 1 and further demonstrate that the second hop is indeed the performance bottleneck for the relaying path.
\item  The accuracy of all developed analytical models are verified via computer-based Monte Carlo simulations. 

\end{itemize}    

\par The remainder of the paper is organised as follows. In Section II, the system model and assumptions are specified. A unif\mbox{}ied framework for \textit{noncoherent} SWIPT relay systems based on TS and PS relaying protocols is developed in Section III. In Section IV, analytical closed-form and approximate expressions for the moments and the MGF of the receive SNR are derived, respectively. The evaluations of ASNR, AoF, outage probability, achievable throughput, and ASER are presented in Section V.  We provide a thorough asymptotic analysis of the considered system model in Section VI. Simulation results are illustrated in Section VII to corroborate the developed analytical models and to investigate the effect of various system parameters on the performance. Some concluding remarks are given in Section VIII. 
 \par \underline{\textit{Notation}:} Bold lower case letters denote vectors and lower case letters denote scalers. A circularly symmetric complex Gaussian random variable $z$ with mean $\mu$ and variance $\sigma^2$ is represented as $\mathcal{CN}(\mu,\sigma^2)$. Also, $\textbf{I}_N$ and $\textbf{i}_{n}$ denote the identity matrix of size $N$ and a column vector with 1 at its $n$-th entry and 0 elsewhere, respectively. Moreover, $(.)^T$, $\mathbb{E}[z]$, $|z|$, $\Re\left\{.\right\}$, and ln(.) stand for the transpose, expectation of the random variable $z$, magnitude of a complex variable $z$, real part, and natural logarithm, respectively.  
\section{System Model}
\label{sec:model}
Consider an AF wireless relay system where a source node, $S$, communicates with a destination node, $D$, via an intermediate relay node, $R$. Let $h_{sr}$ and $h_{rd}$ denote the channel coefficients of the first and the second hops, respectively. The source and the destination nodes are assumed to be energy unconstrained nodes powered by either a battery or a power grid. Furthermore, the source transmits its message using a constant transmit power, $P_s$. On the other hand, the relay node have no dedicated power supply and it harvests energy from the received signal of $S$, which is then used as a transmit power over the second hop. Our study considers two wireless EH protocols, namely, PS and TS, and assumes that all nodes are equipped with a single antenna. We further consider that a direct link does not exist between $S$ and $D$. To avoid interference, data transmission from $S$ and $R$ are performed over orthogonal channels in either frequency or time. For ease of exposition, we consider a time division multiple access (TDMA) based transmission scheme, which was initially proposed for conventional relay systems with \textit{coherent} detection \cite{Nabar} and was later applied to noncoherehnt/differiential systems \cite{Zhao}. All nodes are assumed to operate in half-duplex mode. That is, the signal transmission from $S$ to $D$ is completed in two phases. In \textit{Phase-1}, the source to relay information transmission takes place. During \textit{Phase-2}, $S$ remains silent, whereas $R$ uses the energy harvested during \textit{Phase-1} to amplify the received signal and forward it to $D$. 
\par In our setup, $h_{ij}, ij \in \lbrace{sr,rd\rbrace },$  denotes the small-scale fading coefficient of the channel between transmitter $i$ and receiver $j$, and is modelled as independent complex Gaussian random variable, $h_{ij} \sim \mathcal{C}\mathcal{N}(0,\sigma^2_{ij})$. Moreover, we assume that the instantaneous CSIs of $h_{ij}$ are unknown to any terminal. Furthermore, we consider that all links are subject to large scale fading in which the received power is inversely proportional to $d_{ij}^{−\alpha}$, where $d_{ij}$ is the distance between transmitter $i$ and receiver $j$, and $\alpha > 2$ denotes the path-loss exponent. We also assume that information transfer is performed using \textit{noncoherent} signalings such as $M$-FSK or $M$-DPSK, where $M$ is the constellation size.    
\subsection{Power Splitting-Based Relaying}
\par
\begin{figure}[!t]
\centering
   \includegraphics[width=3.3in]{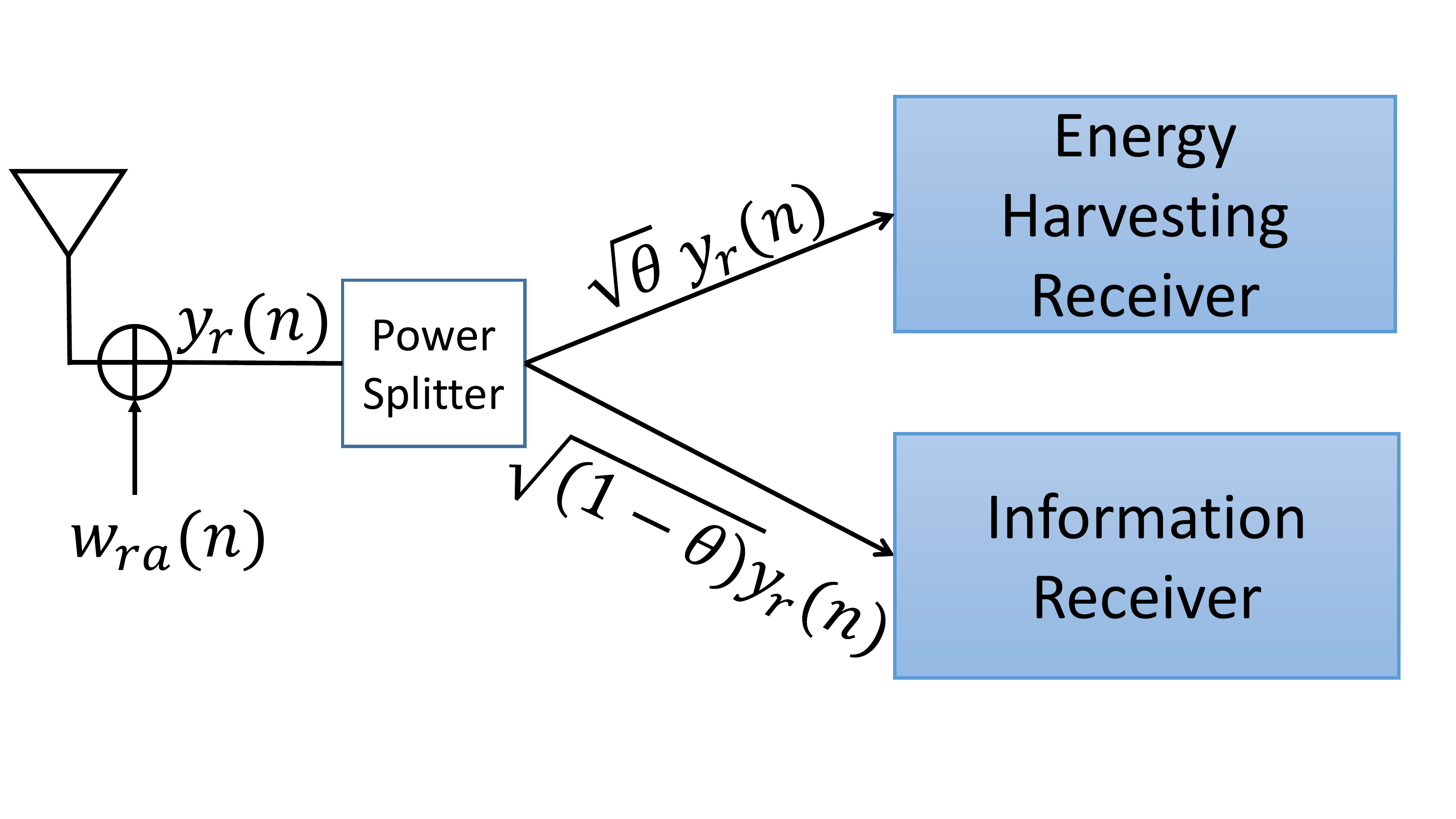}
   \caption{Block diagram of the relay receiver architecture with the PS relaying protocol.}
   \label{f1} 
    \vspace*{-1cm}
\end{figure}

 The communication block diagram of the employed PS relaying protocol for SWIPT is depicted in Fig. \ref{f1}. The total transmission time, $T$, is divided equally into two consecutive phases, each of duration $T/2$ (sec) \cite{Nasir2013,Evans,Liu}. Specifically, during \textit{Phase-1}, the source broadcasts its signal $x_m$, where $\mathbb{E}[|x_m(n)|^2]=1$, with power $P_s$ (watts) according to one of the two following \textit{noncoherent} modulation schemes. 
\subsubsection{PS with $M$-DPSK}
\par  In this case, the information symbols, $v(n)=e^{j2\pi m/M}$ for $m=0,1,...,M-1$, are encoded based on the difference of two consecutive signal phases. Consequently, the source transmitted signal, $x_m, m=0,1,...,M-1$, can be expressed as  
\begin{equation}
x_m(n)=x_m(n-1)v(n), \text{    }n=1,2,...N, 
\end{equation}where the initial reference modulated symbol is $x_m(0)=1$, and $N$ is the frame length \cite{Proakis}. More specifically, in the first time slot, the signal received at $R$ can be expressed as
\begin{align}
y_{r}(n)=\frac{\sqrt{P_s}}{\sqrt{d_{sr}^\alpha}}h_{sr}x_m(n)+w_{ra}(n).
\end{align}Here, $w_{ra}(n)\sim \mathcal{C}\mathcal{N}(0,N_{0_{ra}})$ represents the additive white Gaussian noise (AWGN) associated with the $n$th symbol, accounting for the receive antenna noise at $R$. At the end of \textit{Phase-1}, $R$ splits the received signal into two streams, as shown in Fig. \ref{f1}, one is forwarded to the energy harvester and the other to the information receiver (IR) for information processing. Thus, the signal received at IR of $R$ can be given by
\begin{equation}
\label{YIR}
y_{sr}(n)=\sqrt{\kappa}y_{r}(n)=\frac{\sqrt{\kappa P_s}}{\sqrt{d_{sr}^\alpha}}h_{sr}x_m(n)+\sqrt{\kappa}w_{ra}(n)+w_{rc}(n),
\end{equation} where $\kappa=(1-\theta)$ and $\theta$ stands for the PS ratio at $R$. In this paper, it is considered that $0<\theta<1$, corresponding to a general SWIPT system featuring both wireless information transfer and wireless EH. Furthermore, $w_{rc}(n)\sim \mathcal{C}\mathcal{N}(0, N_{0_{rc}})$ is the AWGN at $R$ due to the RF to baseband signal conversion. 
\par The remaining portion of the received signal at $R$ is forwarded to the energy harvester, i.e. $y_{\text{EH}}(n)=\sqrt{\theta}y_{r}(n)$, hence, the overall energy harvested during \textit{Phase-1} can be expressed as 
\begin{equation}
E_r=\frac{\eta\theta P_s|h_{sr}|^2}{d_{sr}^\alpha}(T/2),
\end{equation} with $0<\eta<1$ denoting the energy conversion efficiency factor. Since $R$ communicates with $D$ for $T/2$ sec., the power available at $R$ at the end of \textit{Phase-1} is $P_r = E_r/(T/2)$. 
\par During \textit{Phase-2} and to simplify the analysis, $R$ uses the whole portion of the harvested energy to amplify the received signal \eqref{YIR} and transmit it to $D$. At the end of \textit{Phase-2}, the destination implementes different decoding to recover the data symbols from the received signal, which can be expressed as
\begin{equation}\label{YD}
y_{rd}(n) = \frac{1}{\sqrt{d_{rd}^\alpha}}h_{rd}s_r(n)+w_{rd}(n),
\end{equation} where $w_{rd}(n)\sim \mathcal{C}\mathcal{N}(0, N_{0_{rd}})$ is the overall AWGN at $D$. In addition, $w_{ra}(n)$,  $w_{rc}(n)$, and  $w_{rd}(n)$ are statistically independent. In (\ref{YD}), $s_r(n)$ represents the signal transmitted from the relay after proper amplification. To ensure that the average transmit power of $R$ is $P_r$, the relay normalises the signal to be transmitted by the variance of (\ref{YIR}) and $s_r(n)$ can be correspondingly, given as 
\begin{equation}\label{XR}
s_r(n)=\sqrt{\frac{P_r }{\frac{1}{d_{sr}^\alpha}\kappa P_s \sigma_{sr}^2+\kappa N_{0_{ra}}+N_{0_{rc}}}}y_{sr}(n).
\end{equation}
We note that, the power scaling factor in Eq. (\ref{XR}), i.e., $(\frac{1}{d_{sr}^\alpha}\kappa P_s\sigma_{sr}^2+\kappa N_{0_{ra}}+N_{0_{rc}})^{1/2}$ ensures that the average transmission power for data relaying in \textit{Phase-2} is fixed to $P_r$ and relies solely on the variance of the channel $h_{sr}$ without
requiring the knowledge of its CSI at the relay terminal \cite{Nasir2013, Liu}. It should be further noted that the EH process at $R$ is independent of the power scaling process and it is assumed that EH is performed instantaneously. We would like to point out that the harvested instantaneous energy depends on the instantaneous power gain, namely, $|h_{sr}|^2$ , but it does not require the knowledge of CSI. The harvested instantaneous energy is simply used as a transmit power in the second phase of transmission. The assumption of instantaneous EH can be justified in the absence of a storage device. Finally, substituting (\ref{YIR}) and (\ref{XR}) in (\ref{YD}) yields 
\begin{equation}\label{YD2}
y_{rd}(n)=\underbrace{\frac{\sqrt{\kappa \eta \theta |h_{sr}|^2}P_s h_{sr}h_{rd}x_m(n)}{\sqrt{d_{sr}^\alpha d_{rd}^\alpha}\sqrt{\kappa P_s  \sigma_{sr}^2+d_{sr}^\alpha N_{0_{sr}}}}}_{\text{Signal Part}}+\underbrace{\frac{\sqrt{\eta \theta P_s |h_{sr}|^2}h_{rd}w_{sr}(n)}{\sqrt{d_{rd}^\alpha}\sqrt{\kappa P_s  \sigma_{sr}^2+d_{sr}^\alpha N_{0_{sr}}}} + w_{rd}(n)}_{\text{Noise Part}},
\end{equation}where $w_{sr}(n)=\sqrt{\kappa} w_{ra}(n)+w_{rc}(n)$ is the overall noise at $R$ with variance $N_{0_{sr}}=\kappa N_{0_{ra}}+N_{0_{rc}}$. Based on (\ref{YD2}), the instantaneous received SNR at $D$, $\gamma_{eq}$, can be given by 
\begin{equation}\label{SNReq PS1}
\gamma_{eq}=\frac{a b \rho_{sr} \rho_{rd}|h_{sr}|^4|h_{rd}|^2}{b \rho_{rd}|h_{sr}|^2|h_{rd}|^2+ a \rho_{sr}\sigma^2_{sr}+1},
\end{equation}
where $a=\frac{\kappa}{d_{sr}^\alpha}$, $b=\frac{\eta \theta}{d_{sr}^\alpha d_{rd}^\alpha}$, $\rho_{sr} = \frac{P_s}{N_{0_{sr}}}$ and $\rho_{rd}=\frac{P_s}{N_{0_{rd}}}$. 
\subsubsection{PS with \textit{noncoherent} $M$-FSK}
\par For \textit{noncoherent} $M$-ary FSK modulation, the source transmits the message, $x_m(n)$, over the $m$th orthogonal carrier chosen from an $M$ set of carriers where $m=0,1,...,M-1$. Since at the destination, the source message is detected based on $M$ received symbols, it is convenient to represent $x_m(n)$ as a column vector $\textbf{i}_{m+1}$ with 1 at its $m$th entry and 0 elsewhere. Moreover, the destination employs \textit{noncoherent} reception using a bank of $M$ \textit{noncoherent} correlators to make a decision as to which of the $M$ symbols was transmitted \cite{Proakis}. The baseband equivalent signal at any receiving terminal, $j$ is denoted by an $M\times 1$ vector $\textbf{y}_{ij}\triangleq\left[y_{ij}(1),...,y_{ij}(M)\right]^T$. As a result, the signal model for PS relaying employing \textit{noncoherent} $M$-FSK can be represented as
\begin{equation}
\textbf{y}_{sr}(n)=\frac{\sqrt{\kappa P_s}}{\sqrt{d_{sr}^\alpha}}h_{sr}\textbf{i}_{m+1}(n)+\sqrt{\kappa}\textbf{w}_{ra}(n)+\textbf{w}_{rc}(n),
\end{equation}
\begin{equation}\label{YRD2}
\textbf{y}_{rd}(n) = \frac{1}{\sqrt{d_{rd}^\alpha}}h_{rd}\textbf{s}_r(n)+\textbf{w}_{rd}(n),
\end{equation}where $\textbf{w}_{ra}(n)\sim \mathcal{C}\mathcal{N}(0,N_{0_{ra}}\textbf{I}_M)$ and $\textbf{w}_{rc}(n)\sim \mathcal{C}\mathcal{N}(0,N_{0_{rc}}\textbf{I}_M)$ are the receive antenna and RF to baseband signal conversion AWGNs at $R$, respectively, and $\textbf{w}_{rd}(n)\sim \mathcal{C}\mathcal{N}(0,N_{0_{rd}}\textbf{I}_M)$ is the overall AWGN due to both the receive antenna and RF to baseband signal conversion at $D$. Furthermore, $\textbf{s}_r(n)$ represents the signal transmitted from $R$ during \textit{Phase}-2 after normalisation, where the amplification gain is chosen to ensure that the average transmission power at $R$ is fixed to $P_r$ and thus, $\textbf{s}_r(n)$ can be expressed as
\begin{equation}\label{SR 2}
\textbf{s}_r(n)=\sqrt{\frac{P_r}{\frac{1}{d_{sr}^\alpha}\kappa P_s \sigma_{sr}^2+M \left[ \kappa N_{0_{ra}}+N_{0_{rc}} \right]}}\textbf{y}_{sr}(n).
\end{equation} By substituting \eqref{SR 2} in \eqref{YRD2}, the signal received at $D$ is given as
\begin{equation}\label{YRD22}
\textbf{y}_{rd}(n)=\underbrace{\frac{\sqrt{\kappa \eta \theta |h_{sr}|^2}P_s h_{sr}h_{rd}\textbf{i}_{m+1}(n)}{\sqrt{d_{sr}^\alpha d_{rd}^\alpha}\sqrt{\kappa P_s \sigma_{sr}^2+M d_{sr}^\alpha N_{0_{sr}}}}}_{\text{Signal Part}}+\underbrace{\frac{\sqrt{\eta \theta P_s |h_{sr}|^2}h_{rd}\textbf{w}_{sr}(n)}{\sqrt{d_{rd}^\alpha}\sqrt{\kappa P_s \sigma_{sr}^2+M d_{sr}^\alpha N_{0_{sr}}}} + \textbf{w}_{rd}(n)}_{\text{Noise Part}},
\end{equation} where $\textbf{w}_{sr}(n)=\sqrt{\kappa} \textbf{w}_{ra}(n)+\textbf{w}_{rc}(n)$ is the overall noise at $R$ with variance $N_{0_{sr}}=\kappa N_{0_{ra}}+N_{0_{rc}}$. Consequently, the instantaneous received SNR at $D$ can be expressed as
\begin{equation}\label{SNReq PS2}
\gamma_{eq}=\frac{a b \rho_{sr} \rho_{rd}|h_{sr}|^4|h_{rd}|^2}{b \rho_{rd}|h_{sr}|^2|h_{rd}|^2+ a \rho_{sr}\sigma^2_{sr}+M},
\end{equation} where $a=\frac{\kappa}{d_{sr}^\alpha}$, $b=\frac{\eta \theta}{d_{sr}^\alpha d_{rd}^\alpha}$, $\rho_{sr} = \frac{P_s}{N_{0_{sr}}}$ and $\rho_{rd}=\frac{P_s}{N_{0_{rd}}}$. In the aforementioned \textit{noncoherent} modulation schemes, it is assumed that all waveforms within one symbol interval are equiprobable and have the same energy $P_s$. 
\subsection{Time Switching-Based Relaying}
\begin{figure*}[!t]
\centering
   \begin{subfigure}[b]{0.5\textwidth}
   \includegraphics[width=\columnwidth]{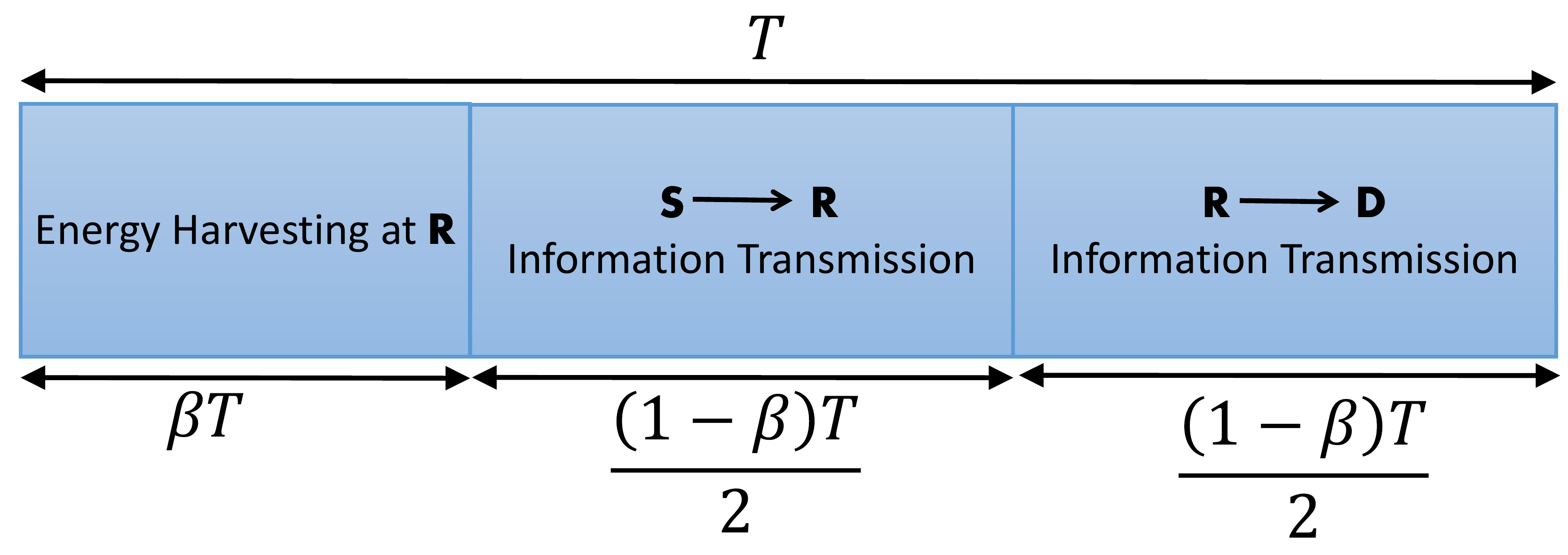}
   \caption{}
   \label{f2_a} 
\end{subfigure}\hfill
\begin{subfigure}[b]{0.5\textwidth}
   \includegraphics[width=\columnwidth]{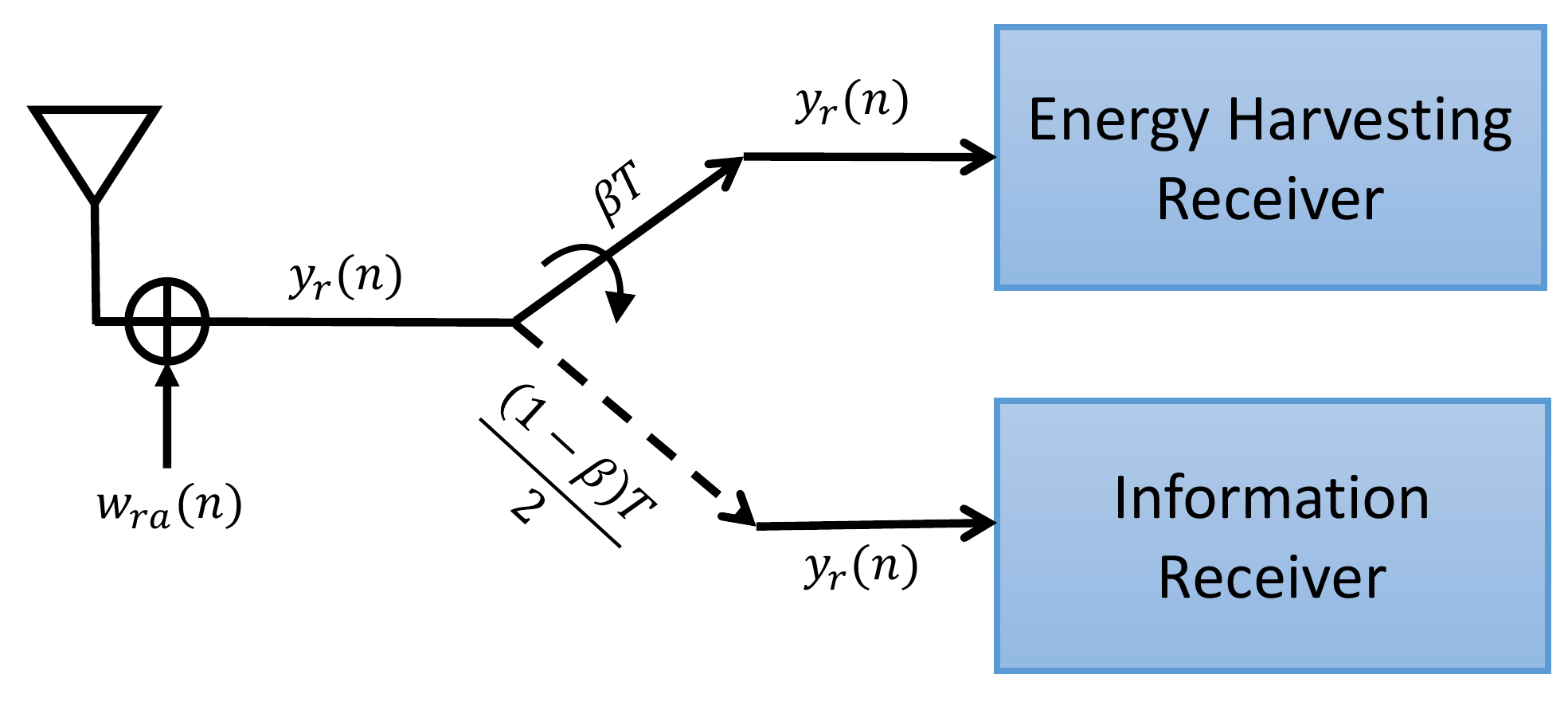}
   \caption{}
   \label{f2_b}
\end{subfigure}
\caption{(a) Illustration of the time frame of SWIPT relaying employing the TS protocol (b) Block diagram of the relay receiver architecture with the TS relaying protocol.}
 \vspace*{-1cm}
 \end{figure*}

The key parameters in the TS relaying protocol for EH and information processing at $R$ are depicted in Fig. \ref{f2_a}. A certain block of information is transmitted from $S$ to $D$ during a total time of $T$ (sec). Under this protocol, $\beta$ is the portion of the total time in which the relay harvests energy from the source signal, where $0<\beta<1$. Following the same setup as in \cite{Nasir2013,Liu} for TS, the information transmission is performed in the remaining block time, i.e., $(1-\beta)T$, such that half of the fraction of the time, $(1-\beta)T/2$, denoted as \textit{Phase}-1, is used for the source to relay transmission, while the remaining fraction of the time, i.e., $(1-\beta)T/2$, denoted as \textit{Phase}-2, is used by the relay to forward the information to the destination as shown in Fig. \ref{f2_b} .  Based on that, the energy harvested at $R$ is given by 
\begin{equation}
E_r=\frac{\eta P_s|h_{sr}|^2}{d_{sr}^\alpha}(\beta T).
\end{equation} Based on the fact that $R$ communicates with $D$ for the time $(1-\beta)T/2$, the transmitted power from the relay node is given by
\begin{equation}\label{Pr TS}
P_r=\frac{E_r}{(1-\beta)T/2}=\frac{2\eta P_s|h_{sr}|^2 \beta}{d_{sr}^\alpha (1-\beta)}.
\end{equation} In the following, we present the details of the signal model when \textit{noncoherent} $M$-DPSK and $M$-FSK are employed with TS-based relaying. 
\subsubsection{TS with $M$-DPSK}
At the end of \textit{Phase}-1, the signal received from $S$ is used solely for information processing at $R$, i.e., no PS is involved and accordingly, the baseband equivalent signal model with $M$-DPSK is given as
\begin{equation}
\label{YIR TS}
y_{sr}(n)=\frac{\sqrt{P_s}}{\sqrt{d_{sr}^\alpha}}h_{sr}x_m(n)+w_{ra}(n)+w_{rc}(n),
\end{equation}
\begin{equation}\label{YD2 TS}
y_{rd}(n) = \frac{1}{\sqrt{d_{rd}^\alpha}}h_{rd}s_r(n)+w_{rd}(n),
\end{equation} where $w_{ra}(n)\sim \mathcal{C}\mathcal{N}(0,N_{0_{ra}})$ and $w_{rc}(n)\sim \mathcal{C}\mathcal{N}(0,N_{0_{rc}})$ are the receive antenna and RF to baseband signal conversion AWGNs at $R$ and $w_{rd}(n)\sim \mathcal{C}\mathcal{N}(0,N_{0_{rd}})$ is the overall AWGN due to the receive antenna and RF to baseband signal conversion at $D$, respectively. To ensure that the power of the transmitted signal from $R$ for TS-based relaying with $M$-DPSK is set to $P_r$, the signal forwarded from the relay node, $s_r(n)$ is given by
\begin{equation}\label{XR TS 1}
s_r(n)=\sqrt{\frac{P_r}{\frac{1}{d_{sr}^\alpha} P_s \sigma_{sr}^2+ N_{0_{ra}}+N_{0_{rc}}}}y_{sr}(n).
\end{equation} By setting $w_{sr}(n) = w_{ra}(n)+w_{rc}(n)$, $N_{0_{sr}}=N_{0_{ra}}+N_{0_{rc}}$, and upon substituting \eqref{XR TS 1} and \eqref{Pr TS} into \eqref{YD2 TS}, the received signal at $D$, $y_{rd}(n)$, in terms of $P_s$, $\eta$, $\beta$, $d_{sr}^\alpha$, and $d_{rd}^\alpha$, can be rewritten as 
\begin{equation}\label{YD22 TS}
y_{rd}(n)=\underbrace{\frac{\sqrt{2 \eta \beta |h_{sr}|^2}P_s h_{sr}h_{rd}x_m(n)}{\sqrt{(1-\beta )d_{sr}^\alpha d_{rd}^\alpha}\sqrt{P_s \sigma_{sr}^2+d_{sr}^\alpha N_{0_{sr}}}}}_{\text{Signal Part}}+\underbrace{\frac{\sqrt{2\eta \beta P_s |h_{sr}|^2}h_{rd}w_{sr}(n)}{\sqrt{(1-\beta )d_{rd}^\alpha}\sqrt{P_s \sigma_{sr}^2+d_{sr}^\alpha N_{0_{sr}}}} + w_{rd}(n)}_{\text{Noise Part}}.
\end{equation} From (\ref{YD22 TS}), the instantaneous received SNR of the relay path, $\gamma_{\text{eq}}$, can be expressed as \eqref{SNReq PS1} with $a=\frac{1}{d_{sr}^\alpha}$, $b = \frac{2\eta \beta}{d_{sr}^\alpha d_{rd}^\alpha (1-\beta )}$, $\rho_{sr} = \frac{P_s}{N_{0_{sr}}}$, and $\rho_{rd} = \frac{P_s}{N_{0_{rd}}}$. 
\subsubsection{TS with \textit{noncoherent} $M$-FSK}
When \textit{noncoherent} $M$-FSK signaling is employed, the signals received at $R$ at the end of \textit{Phase}-1 and at $D$ at the end of \textit{Phase}-2 can be respectively modeled as 
\begin{equation}
\textbf{y}_{sr}(n)=\frac{\sqrt{ P_s}}{\sqrt{d_{sr}^\alpha}}h_{sr}\textbf{i}_{m+1}(n)+\textbf{w}_{ra}(n)+\textbf{w}_{rc}(n),
\end{equation}
\begin{equation}\label{YRD2 TS}
\textbf{y}_{rd}(n) = \frac{1}{\sqrt{d_{rd}^\alpha}}h_{rd}\textbf{s}_r(n)+\textbf{w}_{rd}(n),
\end{equation}where $\textbf{w}_{ra}(n)\sim \mathcal{C}\mathcal{N}(0,N_{0_{ra}}\textbf{I}_M)$ and $\textbf{w}_{rc}(n)\sim \mathcal{C}\mathcal{N}(0,N_{0_{rc}}\textbf{I}_M)$ are the receive antenna and RF to baseband signal conversion AWGNs at $R$, respectively, and $\textbf{w}_{rd}(n)\sim \mathcal{C}\mathcal{N}(0,N_{0_{rd}}\textbf{I}_M)$ is the overall AWGNs due to the receive antenna and RF to baseband signal conversion at $D$. The signal received at $R$ in \textit{Phase}-1 is amplified or scaled to meet an average power constraint $P_r$ and thus, the transmitted signal, $\textbf{s}_r(n)$, is specified as
\begin{equation}\label{SR2 TS}
\textbf{s}_r(n)=\sqrt{\frac{P_r}{\frac{1}{d_{sr}^\alpha} P_s \sigma_{sr}^2+M \left[  N_{0_{ra}}+N_{0_{rc}} \right]}}\textbf{y}_{sr}(n).
\end{equation} Combining \eqref{Pr TS} and \eqref{SR2 TS}, followed by a substitution of \eqref{SR2 TS} back into \eqref{YRD2 TS}, yields the signal received at $D$ as
\begin{equation}\label{YD22 TS2}
y_{rd}(n)=\underbrace{\frac{\sqrt{2 \eta \beta |h_{sr}|^2}P_s h_{sr}h_{rd}\textbf{i}_{m+1}(n)}{\sqrt{(1-\beta )d_{sr}^\alpha d_{rd}^\alpha}\sqrt{P_s \sigma_{sr}^2+M d_{sr}^\alpha N_{0_{sr}}}}}_{\text{Signal Part}}+\underbrace{\frac{\sqrt{2\eta \beta P_s |h_{sr}|^2}h_{rd}\textbf{w}_{sr}(n)}{\sqrt{(1-\beta )d_{rd}^\alpha}\sqrt{P_s \sigma_{sr}^2+M d_{sr}^\alpha N_{0_{sr}}}} + \textbf{w}_{rd}(n)}_{\text{Noise Part}},
\end{equation} where $\textbf{w}_{sr}(n)= \textbf{w}_{ra}(n)+\textbf{w}_{rc}(n)$ is the overall noise at $R$ with variance $N_{0_{sr}}= N_{0_{ra}}+N_{0_{rc}}$. Based on \eqref{YD22 TS2}, the instantaneous received SNR of the relay link, when TS relaying is employed with \textit{noncoherent} $M$-FSK, is expressed as \eqref{SNReq PS2} with $a=\frac{1}{d_{sr}^\alpha}$, $b = \frac{2\eta \beta}{d_{sr}^\alpha d_{rd}^\alpha (1-\beta )}$, $\rho_{sr} = \frac{P_s}{N_{0_{sr}}}$, and $\rho_{rd} = \frac{P_s}{N_{0_{rd}}}$.  
\section {A Unif\mbox{}ied System Model for \textit{NONCOHERENT} Modulation in SWIPT}
\par In the previous section, we developed four distinct system models with a set of different parameters to feature four possible scenarios of \textit{noncoherent} SWIPT relay systems. This includes PS-based and TS-based relaying protocols with \textit{noncoherent} $M$-DPSK and $M$-FSK. To simplify the ensuing analysis, we present a unifying framework for the four instantaneous received SNR expressions as follows
\begin{equation}\label{uniSNR}
\gamma_{eq}=\frac{\hat{a}\hat{b}\gamma_{sr}^2 \gamma_{rd}}{\hat{b}\gamma_{sr}\gamma_{rd} + \hat{a}\sigma^2_{sr}+ \Psi}, 
\end{equation}where $\gamma_{sr}=|h_{sr}|^2 $ and $\gamma_{rd}=|h_{rd}|^2 $. Also,  $\hat{a}=\frac{\kappa}{d_{sr}^\alpha}\rho_{sr}$ for the PS-based relaying and $\hat{a}=\frac{1}{d_{sr}^\alpha}\rho_{sr}$ for TS-based relaying. Moreover, $\hat{b}=\frac{\eta \theta}{d_{sr}^\alpha d_{rd}^\alpha}\rho_{rd}$ for PS-based relaying and $\hat{b}=\frac{2\eta \beta}{d_{sr}^\alpha d_{rd}^\alpha (1-\beta )}\rho_{rd}$ for TS-based relaying. It is worth noting that for both TS and PS relaying schemes, $\rho_{sr}$ and $\rho_{rd}$ are set to $\rho_{sr} = P_s/N_{0_{sr}}$ and $\rho_{rd} = P_s/N_{0_{rd}}$, respectively. The effective noise variance at the relay terminal, $N_{0_{sr}}$, accounting for the two protocols, can be given as $N_{0_{sr}}=\kappa N_{0_{ra}}+N_{0_{rc}}$ for PS-based relaying and $N_{0_{sr}}=N_{0_{ra}}+N_{0_{rc}}$ for TS-based relaying. Furthermore, $\Psi$ is set to 1 for $M$-DPSK, and $M$ for $M$-FSK. Note that the unif\mbox{}ied received SNR expression in \eqref{uniSNR} can represent one of the four possible scenarios, i.e., PS-based relaying with either $M$-DPSK or $M$-FSK, and TS-based relaying with either $M$-DPSK or $M$-FSK. 
\section{The Moments and MGF of the Receive SNR} 
\par In this section, we derive an exact unif\mbox{}ied closed-form expression for the moments of the receive SNR, $\gamma_{eq}$ in \eqref{uniSNR}, and then employ it to approximate its MGF .
\subsection{Moments of the Receive SNR}
Since deriving the PDF and CDF of the receive SNR expression in \eqref{uniSNR} is a challenging task, due to the term $\gamma_{sr}^2=|h_{sr}|^4$ in the numerator, we alternatively focus on deriving its moments. The moments, specified as $\mu_n=\mathbb{E}[\gamma^n_{eq}]$, can be derived in the following theorem.
\begin{theorem}
Assuming that both $S\to R$ and $R\to D$ links undergo independent and identically distributed (i.i.d) Rayleigh fading conditions, the $n^\text{th}$-order moment of the instantaneous end-to-end SNR of dual-hop AF relaying in \textit{noncoherent} SWIPT systems is given by
\begin{equation}\label{moment final}
\mu_n=\mathbb{E}[\gamma^n_{eq}]=\frac{\hat{a}^n C {(\lambda_{{sr}}})^{n-1}}{\hat{b}\lambda_{{rd}} \Gamma(n)} G^{3, 1}_{1, 3}\left[\frac{\hat{b}\lambda_{{sr}}\lambda_{{rd}}}{C} \  \big\vert \  {1, 1-n, 2 \atop n+1}\right],
\end{equation}where $ G^{.,.}_{.,.}[.\vert .]$ is the Meijer G-function as defined in \cite[Eq. (8.2.1.1)]{Prudnikov}.
\end{theorem}
\begin{IEEEproof}
See Appendix \ref{Appendix A}.
\end{IEEEproof}
\par To the best of our knowledge, this result is novel. It is worth noting that \eqref{moment final} is simple and incorporates the Meijer G-function which is a standard built-in function in most of the well-known mathematical software packages, such as MATLAB, MAPLE, and MATHEMATICA, and can therefore, be easily and efficiently evaluated. Theorem 1 turns out to be useful beyond the scope of this paper. Knowing \eqref{moment final}, one can establish the MGF, AoF, ASNR, outage probability, achievable throughput, and ASER of the system, as will be presented in the subsequent section. Nonetheless, the result in \eqref{moment final} can be further applied to study other metrics, such as the kurtosis and the skewness that characterise the distribution of the receive SNR and the ergodic capacity of the system.
\subsection{MGF of the Receive SNR} \label{MGFPADE}
By definition, the MGF of $\gamma_{eq}$ is $
\mathcal{M}_{\gamma_{eq}}(s)=\mathbb{E}\left[e^{-s\gamma_{eq}}\right]$ and can be represented as a formal power series (e.g., Taylor) as 
\begin{equation}\label{MGF series}
\mathcal{M}_{\gamma_{eq}}(s)=\sum_{n=0}^{\infty}\frac{(-1)^n}{n!}\mathbb{E}\left[\gamma_{eq}^n\right]s^n=\sum_{n=0}^{\infty}\frac{(-1)^n}{n!}\mu_n s^n.
\end{equation} Despite the fact that the moments of all orders, $\mu_n$, can be computed in closed-form using the analysis of the preceding subsection, in many cases, we cannot conclude where and whether the power series in \eqref{MGF series} is convergent or not. Hence, in practice, only a finite number $W$ can be used, truncating the series \eqref{MGF series} as
\begin{equation}\label{MGF trun}
\mathcal{M}_{\gamma_{eq}}(s)=\sum_{n=0}^{W}\frac{(-1)^n}{n!}\mu_n s^n+\mathcal{O}(s^{W+1}),
\end{equation} with $\mathcal{O}(s^{W+1})$ denoting the terms of order higher than $W$ after the truncation. Having \eqref{MGF trun} in hand, one has to obtain the best approximation to the unknown underlying function $\mathcal{M}_{\gamma_{eq}}(s)$ by evaluating only a finite number of the moments. It is demonstrated in \cite{Baker,Amindavar} that the series, presented in \eqref{MGF series}, can be efficiently and accurately approximated using the PA method. PA is a well-known method that is applied to approximate infinite power series that are either not guaranteed to converge, converge very slowly or for which a limited number of coefficients is known \cite{Baker}. The approximation is given in terms of a simple rational function of arbitrary order $X$ for the numerator and arbitrary order $Y$ for the denominator, and whose power series expansion agrees with the $W$th-order ($W=X+Y$) power expansion of $\mathcal{M}_{\gamma_{eq}}(s)$ \cite{Baker}. Consequently, the rational function
\begin{equation}\label{MGF}
P_{[X/Y]}(s)\triangleq\frac{\sum_{i=0}^{X}x_i s^i}{1+\sum_{i=1}^{Y}y_i s^i}
\end{equation} is said to be a PA to the series \eqref{MGF series}, if
\begin{equation}\label{pade1}
P_{[X/Y]}(s)\cong\sum_{n=0}^{W}\frac{(-1)^n}{n!}\mu_n s^n+\mathcal{O}(s^{W+1}).
\end{equation} It is straight forward to see from \eqref{MGF} that the moments $\mu_n, n=1,..., W$ need to be evaluated to construct the approximant, $P_{[X/Y]}(s)$.  Also, the coefficients $x_i$ and $y_i$ can be easily obtained by matching the coefficients of like powers on both sides. Several issues concerning approximants and the method to determine their coefficients are included in \cite{Baker,Amindavar}. Pad$\acute{\text{e}}$ approximants are available as built-in functions in off-the-shelf mathematical software packages, such as MATLAB, MAPLE, and MATHEMATICA. In this work, we apply the subdiagonal PA, $P_{[X/X+1]}(s)$, to approximate $\mathcal{M}_{\gamma_{eq}}(s)$, since it is only for such order of approximants that the convergence rate and the uniqueness can be assured \cite{Amindavar}.  
\section {Performance Analysis}
In this section we exploit the mathematical tools presented in the previous section to derive a number of important performance measures for SWIPT relay networks employing \textit{noncoherent} modulation.
\vspace*{-0.3cm}
\subsection{Receive Average SNR (ASNR)}
The ASNR is the most common and well understood performance measure characterising a digital communication system, owing to its ease of evaluation. It also serves as an excellent indicator of the overall fidelity of the system \cite{Simon}. The ASNR corresponds to the first moment, $\mu_1=\mathbb{E}[\gamma_{eq}]$, which can be computed by setting $n=1$ in \eqref{moment final}.
\subsection{Amount of Fading (AoF)}
\par For the instantaneous receive SNR, $\gamma_{eq}$, the AoF is defined as \cite{Simon}
\begin{equation}\label{AOF}
AoF_{\gamma_{eq}}=\frac{\mathbb{E}[\gamma^2_{eq}]}{(\mathbb{E}[\gamma_{eq}])^2}-1,
\end{equation} which is obtained by substituting \eqref{moment final} in \eqref{AOF}.
\subsection{Outage Probability}
The outage probability $P_{{out}}$ is another standard performance criterion of wireless systems operating over fading channels. It is defined as the probability that the instantaneous SNR $\gamma_{eq}$ of the $S\to R\to D$ at $D$ falls below a certain predefined threshold $\gamma_{th}$, namely, 
\begin{equation}\label{OP}
P_{{out}}\triangleq F_{\gamma_{eq}}(\gamma_{th})=\text{Pr}(\gamma_{eq}\leq\gamma_{th}),
\end{equation} where $\gamma_{th}=2^{R_T}-1$, $R_T$ is the transmission rate, and $F_{\gamma_{eq}}(.)$ is the CDF of $\gamma_{eq}$, and can be evaluated by \cite{Young-Chai}
\begin{equation}\label{Pout}
P_{{out}}\triangleq F_{\gamma_{eq}}(\gamma_{th})=\left[\mathcal{L}^{-1}\left(\frac{\mathcal{M}_{\gamma_{eq}}(s)}{s}\right)\right]_{s=\gamma_{th}},
\end{equation}where $\mathcal{M}_{\gamma_{eq}}(s)$ is the MGF expression pertaining to $\gamma_{sr}$, and $\mathcal{L}^{-1}(.)$ denotes the inverse Laplace transform. Herein, to evaluate the outage probability, we follow the well-known MGF approach which is achieved by substituting the MGF expression given in \eqref{MGF} into \eqref{Pout}, and applying the accurate Euler numerical technique for the inversion of the Laplace transform presented in \cite{Young-Chai}. Following the steps illustrated therein, the outage probability of the $S\to R\to D$ link of \textit{noncohernt} SWIPT systems can be calculated according to  
\begin{equation}\label{pout1}
P_{{out}}=\frac{2^{-Q}e^{A/2}}{\gamma_{th}} \sum_{q=0}^Q {Q\choose q}\sum_{n=0}^{N+q}\frac{(-1)^n}{\beta_n}\Re\left\{\frac{\mathcal{M}_{\gamma_{eq}}(\frac{A+2\pi jn}{2\gamma_{th}})}{\frac{A+2\pi jn}{2\gamma_{th}}}\right\}+E(A,N,Q),
\end{equation} with $\beta_n=2$ when $n=0$ and $\beta_n=1$ when $n=1,2,...,N$, and $A, Q$ and $N$ denote the truncation parameters. Furthermore, $E(A,N,Q)$ denotes the overall discretisation and truncation error term which can be approximately bounded by \cite{Young-Chai}
\begin{equation}
|E(A,N,Q)|\simeq\frac{e^{-A}}{1-e^{-A}}+\big\vert\frac{2^{-Q}e^{A/2}}{\gamma_{th}} \sum_{q=0}^Q (-1)^{N+1+q}{Q\choose q}\Re\left\{\frac{\mathcal{M}_{\gamma_{eq}}(\frac{A+2\pi j(N+q+1)}{2\gamma_{th}})}{\frac{A+2\pi j(N+q+1)}{2\gamma_{th}}}\right\}\big\vert.
\end{equation}

\subsection {Throughput Analysis}
The throughput, $\tau$, measures the rate of successful information decoding at the destination node, given a fixed transmission rate $R_T$, and is defined for the TS and PS relaying protocols as
\begin{equation}\label{TP1}
\tau_{TS}=\frac{(1-P_{out})R_T(1-\beta)}{2}
\end{equation}and 
\begin{equation}\label{TP2}
\tau_{PS}=\frac{(1-P_{out})R_T}{2},
\end{equation}respectively, \cite{Nasir2013}. 
\vspace*{-0.3cm}
\subsection{Average Symbol Error Rate (ASER)}
The aim of this subsection is to analyse the final performance criterion of the underlying SWIPT relaying scheme. By exploiting the PA method described in Section \ref{MGFPADE}, and the well-known MGF-based unif\mbox{}ied approach to the ASER analysis for \textit{noncoherent} and differential modulation over fading channels \cite{Simon}, the ASER of arbitrary $M$-ary \textit{noncoherent} modulation schemes can be readily evaluated as follows.
\par The unconditional ASER of the proposed system under $M$-ary FSK modulation scheme is given by \cite{Alouini}
\begin{align}\label{ber1 fsk}
\overline{P}_{se_{MFSK}}=\sum_{m=1}^{M-1}(-1)^{m+1}{M-1\choose m}\frac{1}{m+1}\mathcal{M}_{\gamma_{eq}}\left(\frac{m}{m+1}\right).
\end{align}
When $M$-ary DPSK is employed for SWIPT relaying, the ASER is expressed as \cite{Alouini}
\begin{equation}\label{ber1 DPSK}
\overline{P}_{se_{MDPSK}}=\frac{1}{\pi}\int_{0}^{(M-1)\pi /M}\mathcal{M}_{\gamma_{eq}}\left(\frac{g}{1+\sqrt{1-g}\text{cos}\phi}\right)d\phi,
\end{equation}where $g \triangleq\text{sin}^2(\pi/M)$. 
\par Although a closed-form expression cannot be obtained, \eqref{ber1 DPSK} can be easily computed numerically since it present a finite-range summation and integral involving integrands composed of closed-form approximations. By substituting \eqref{MGF} in \eqref{ber1 fsk} and \eqref{ber1 DPSK}, the ASER performance of \textit{noncoherent} $M$-FSK and $M$-DPSK modulation schemes can be evaluated for the desired relaying protocol. 
\vspace*{-0.3cm}
\section{Asymptotic Analysis}
Although the expressions in \eqref{pout1}, \eqref{ber1 fsk}, and \eqref{ber1 DPSK} are based on a very tight approximation of the MGF of the receive SNR, they do not provide useful insights into the system performance. Consequently, in this section, we derive further asymptotic expressions which can be used to provide deep insights into the system performance of the considered system model. 
\par Specifically, considering \eqref{uniSNR}, we can alternatively write it as
\begin{equation}\label{highSNR}
\gamma_{eq}=\frac{\gamma_{\text{hop}_1}\gamma_{\text{hop}_2}}{\gamma_{\text{hop}_2} + \overline{\gamma}_{\text{hop}_1}+ \Psi}, 
\end{equation} where, $\gamma_{\text{hop}_1} =\hat{a}\gamma_{sr}$ and $\gamma_{\text{hop}_2}=\hat{b}\gamma_{sr}\gamma_{rd}$, and $\overline{\gamma}_{\text{hop}_1}=\hat{a}\sigma^2_{sr}$. Examining \eqref{highSNR}, we note that two SNRs, namely, $\gamma_{\text{hop}_1}$ and $\gamma_{\text{hop}_2}$, parameterise the performance. Allowing the two SNRs to become large does not simplify the analysis further; therefore, we allow only $\gamma_{\text{hop}_1}$ to approach infinity, which could be justified when the relay is placed very close to the source. 
In particular, as $\gamma_{\text{hop}_1}\to\infty^+$, \eqref{highSNR} can be expressed as 
\begin{equation}\label{asymSNR}
\gamma_{eq}\cong\hat{b}\gamma_{sr}^2\gamma_{rd},
\end{equation} which follows by setting $\sigma^2_{sr}=1$ for a Rayleigh fading channel. 
\par In the following, we derive asymptotic expressions for the outage probability and the ASER of the system as follows.
\subsection{Asymptotic Outage Probability}
The asymptotic outage probability of \eqref{pout1} is given in the following proposition.
\begin{prop}
The asymptotic outage probability of a \textit{noncoherent} dual-hop SWIPT AF relay network, when the average SNR of the first hop; $S\to R$, approaches $\infty^+$, can be given as
\begin{equation}\label{OPhighSR}
P_{{out 1}}^\infty=F^\infty_{\gamma_{eq}}(\gamma)\cong 1-\frac{1}{\sqrt{\pi}}G^{0, 3}_{3, 0}\left[\frac{4\hat{b}\lambda_{{sr}}^2\lambda_{{rd}}}{\gamma_{th}} \  \big\vert \  {0, 0.5, 1 \atop -}\right]
\end{equation}
\end{prop} 
\begin{IEEEproof}
See Appendix \ref{Appendix B}
\end{IEEEproof}
\par By knowing the mathematical behavior of the Meijer G-function and recalling that $\hat{b}=\frac{\eta \theta}{d_{sr}^\alpha d_{rd}^\alpha}\rho_{rd}$ for PS-based relaying and $\hat{b}=\frac{2\eta \beta}{d_{sr}^\alpha d_{rd}^\alpha (1-\beta )}\rho_{rd}$ for TS-based relaying, where $\rho_{rd} = \frac{P_s}{N_{0_{rd}}}$, we can note from \textit{Proposition} 1 that as $\hat{b}$ increases, corresponding to increasing the average SNR of the second hop, the second term in \eqref{OPhighSR} increases and the outage probability decreases. This suggests that when the average SNR of the first hop is very high, the outage probability of the system decreases regardless of the PS or TS  ratios, i.e., $\theta$ or $\beta$ in PS-based or TS-based relaying, respectively, This will be confirmed in the numerical and simulation results shown in Section VII.   
\par \underline{\textit{Special Case:}} when $\hat{b}\to\infty^+$, corresponding to the case when the average SNR of the second hop; $R\to D$, grows large, we apply the asymptotic representation of the Meijer G-function given in \cite[Eq. (41)]{Ansari} to derive the outage probability in terms of basic elementary functions as
\begin{equation}\label{OPhighSRhighRD}
P_{{out 2}}^\infty=F^\infty_{\gamma_{eq}}(\gamma_{th})\cong 1-\frac{1}{\sqrt{\pi}}\left[\underbrace{\frac{1}{\Omega\hat{b}}\Gamma(-0.5)}_{t_1}+\underbrace{\frac{1}{\sqrt{\Omega\hat{b}}}\Gamma(0.5)\Gamma(-0.5)}_{t_2}+\Gamma(0.5)\right],
\end{equation}where $\Omega=(4\lambda_{{sr}}^2\lambda_{{rd}})/\gamma_{th}$. It is straightforward to notice from \eqref{OPhighSRhighRD} that as $\hat{b}\to\infty^+$  the term $t_2$ dominates over $t_1$. In other words, the convergence of $t_2$ to zero is slower compared to $t_1$, and accordingly, $P_{{out 1,2}}^\infty$ can be reduced to 
\begin{equation}\label{dominantOP}
P_{{out 2}}^\infty=F^\infty_{\gamma_{eq}}(\gamma_{th})\cong -\frac{1}{\sqrt{\pi\Omega\hat{b}}}\Gamma(0.5)\Gamma(-0.5)
\end{equation} 

We focus now on investigating the diversity order to gain further insight into the impact of EH on \textit{noncoherent} relaying systems. Setting $SNR=\hat{b}$, a careful examination of \eqref{dominantOP} reveals that the outage probability of the considered SWIPT relay system for a given rate, $R_T$, for both PS and TS relaying schemes, behaves as $P_{{out 2}}^\infty\propto SNR^{\text{ }-\frac{1}{2}} $  at high SNR, which gives rise to a diversity order $d = 0.5$ as 
\begin{equation}
d=\underset{SNR\to\infty}{\text{lim}}\left(-\frac{\text{log}P_{{out}}}{\text{log}SNR}\right)=-\frac{\text{log}P_{{out 2}}^\infty}{\text{log}SNR}=0.5
\end{equation}It is therefore easy to see from the previous discussion that the obtained diversity order for the considered \textit{noncoherent} SWIPT system is less than 1, while for its \textit{coherent} counterpart it was shown in \cite{Lee} that it is equal to 1. It can also be concluded that the system performance is limited by the second hop; $R \to D$, which is the weaker hop, since it is subject to the cascaded fading effect, resulting from EH at the relay node.

\subsection{Asymptotic ASER}
In order to derive the asymptotic ASER, we utilise the asymptotic CDF, i.e., $F^\infty_{\gamma_{eq}}(\gamma_{th})$ obtained in \eqref{OPhighSR} to derive the asymptotic MGF in the following proposition.
\begin{prop} The asymptotic MGF of the approximate receive SNR of a \textit{noncoherent} dual-hop SWIPT AF relay network when the average SNR of the first hop, $S\to R$, approaches $\infty^+$, can be given as
\begin{equation}\label{SERhighSR}
\mathcal{M}_{\gamma_{eq}1}^\infty(s)\cong 1-\frac{1}{\sqrt{\pi}}G^{1, 3}_{3, 1}\left[4\hat{b}\lambda_{{sr}}^2\lambda_{{rd}}s \  \big\vert \  {0, 0.5, 1 \atop 1}\right].
\end{equation}
\end{prop}
\begin{IEEEproof}
The proof follows by recalling the definition of the MGF, $\mathcal{M}_{\gamma_{eq}}(s)=\mathbb{E}[e^{-s\gamma_{eq}}]$, then using integration by parts to express the MGF in terms of the CDF as
\begin{equation}\label{MGFF}
\mathcal{M}_{\gamma_{eq}}(s)=s\int_0^\infty e^{-s\gamma}F_{\gamma_{eq}}(\gamma)d\gamma.
\end{equation}By substituting the asymptotic CDF of \eqref{OPhighSR} into \eqref{MGFF}, then applying the transformation \cite[Eq. (8.2.2.14)]{Prudnikov}, followed by some mathematical manipulations and finally with the aid of \cite[Eq. (7.813.2)]{Rizhik}, the desired result is obtained.
\end{IEEEproof}
\par \underline{\textit{Special Case:}} when $\hat{b}\to\infty^+$, corresponding to the scenario when the average SNR of the second hop; $R\to D$, grows large, we apply the asymptotic representation of the Meijer G-function given in \cite[Eq. (41)]{Ansari} to derive the MGF in terms of basic elementary functions as
\begin{equation}\label{SERhighSRhighRD}
\mathcal{M}_{\gamma_{eq}2}^\infty(s)\cong 1-\frac{1}{\sqrt{\pi}}\left[\frac{1}{\xi\hat{b}s}\Gamma(-0.5)+\frac{1}{\sqrt{\xi\hat{b}s}}\Gamma(0.5)\Gamma(-0.5)\Gamma(1.5)+\Gamma(0.5)\right],
\end{equation}where $\xi=4\lambda_{{sr}}^2\lambda_{{rd}}$. Similar to \eqref{OPhighSRhighRD}, \eqref{SERhighSRhighRD} can be further reduced to 
\begin{equation}\label{dominantMGF}
\mathcal{M}_{\gamma_{eq}2}^\infty(s)\cong-\frac{1}{\sqrt{\pi\xi\hat{b}s}}\Gamma(0.5)\Gamma(-0.5)\Gamma(1.5).
\end{equation} Capitalising on the previously derived asymptotic MGF expressions, the evaluation of the ASER of $M$-ary FSK and $M$-ary DPSK is discussed in the following Corollary. 
\begin{coro}
The asymptotic ASER of $M$-ary FSK and $M$-ary DPSK of \textit{noncoherent} SWIPT relaying systems can be evaluated by substituting \eqref{SERhighSR} or \eqref{dominantMGF} in \eqref{ber1 fsk} and \eqref{ber1 DPSK}, respectively. For the special cases of binary FSK (BFSK) and BDPSK, the asymptotic ASERs can be obtained in closed-forms by substituting either \eqref{SERhighSR} or \eqref{dominantMGF} in $\overline{P}^\infty_{se_{BFSK}}\cong 0.5\mathcal{M}^\infty_{\gamma_{eq}}(0.5)$ and $\overline{P}^\infty_{se_{BDPSK}}\cong0.5\mathcal{M}^\infty_{\gamma_{eq}}(1)$, respectively \cite{Alouini}.  
\end{coro}
\textit{Proposition} 2 and \textit{Corollary} 1 reveals that as $\hat{b}$ increases, corresponding to increasing the average SNR of the second hop, the second term in \eqref{SERhighSR} increases and $\overline{P}^\infty_{se_{BFSK}}$ and $\overline{P}^\infty_{se_{BDPSK}}$ decrease. This suggests that when the average SNR of the first hop is very high, the ASER of the system decreases regardless of the PS or TS ratios, i.e., $\theta$ or $\beta$ in PS-based or TS-based relaying, respectively.
\vspace*{-0.3cm}

\section{Numerical and Simulation Results}\label{sec:results}
In this section, we provide numerical and simulation results to illustrate and  validate the accuracy of the proposed analytical framework. 
\par Unless otherwise stated, we set the source transmission rate to $R_T=3$ bits/sec/Hz, the EH efficiency $\eta=1$, the source transmission power $P_s=1$ Joules/sec and path loss $\alpha=2.7$ \cite{Nasir2013}. We assume that $N_{0}\triangleq N_{0_{sr}}=N_{0_{rd}}$ and for simplicity, we assume that the antenna noise and information receiver circuit noise at the relay node have equal variances, i.e., $N_{0_{ra}}=N_{0_{rc}}\triangleq N_0/2$. Since the only energy supplied to the whole network is the transmit power, $P_s$, applied to the source, the performance of the whole system is parametrised by SNR $\triangleq P_s/N_0$. Furthermore, the distances $d_{sr}$ and $d_{rd}$ are normalised to unity. Additionally, the mean values, $\lambda_{{sr}}$ and $\lambda_{{rd}}$, of the exponential random variables $\gamma_{sr}$ and $\gamma_{rd}$, respectively, are set to 1. 
\par Following the parameter values used in \cite{Young-Chai}, in our simulation results, we consider $A=23$ and parameters $Q$ and $N$ are set to 15 and 21, respectively, so as to ensure a discretisation error less that $10^{-10}$.  However, the overall resulting error is negligible compared to the actual outage probability value.
\par
\begin{figure}[!t]
\centering
   \includegraphics[width=3.3in]{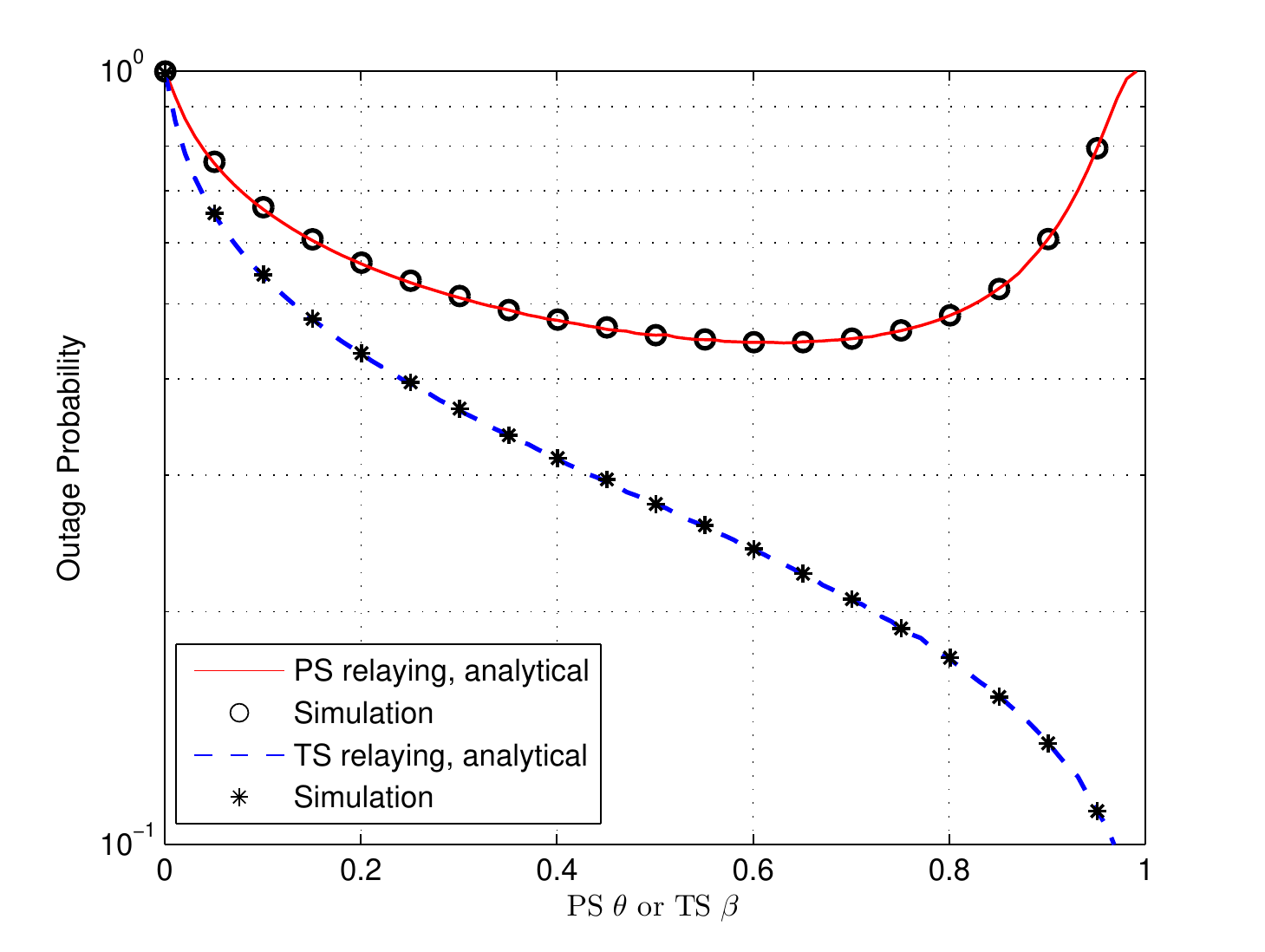}
   \caption{Outage probability, $P_{out}$, with respect to $\beta$ for the TS protocol and $\theta$ for the PS protocol for SNR = 20 dB. $\eta = 1$, $P_s=1$, and $d_{sr}=d_{rd}=1$.}
   \label{f5} 
   \vspace*{-1cm}
\end{figure}

\begin{figure}[!t]
\centering
   \includegraphics[width=3.3in]{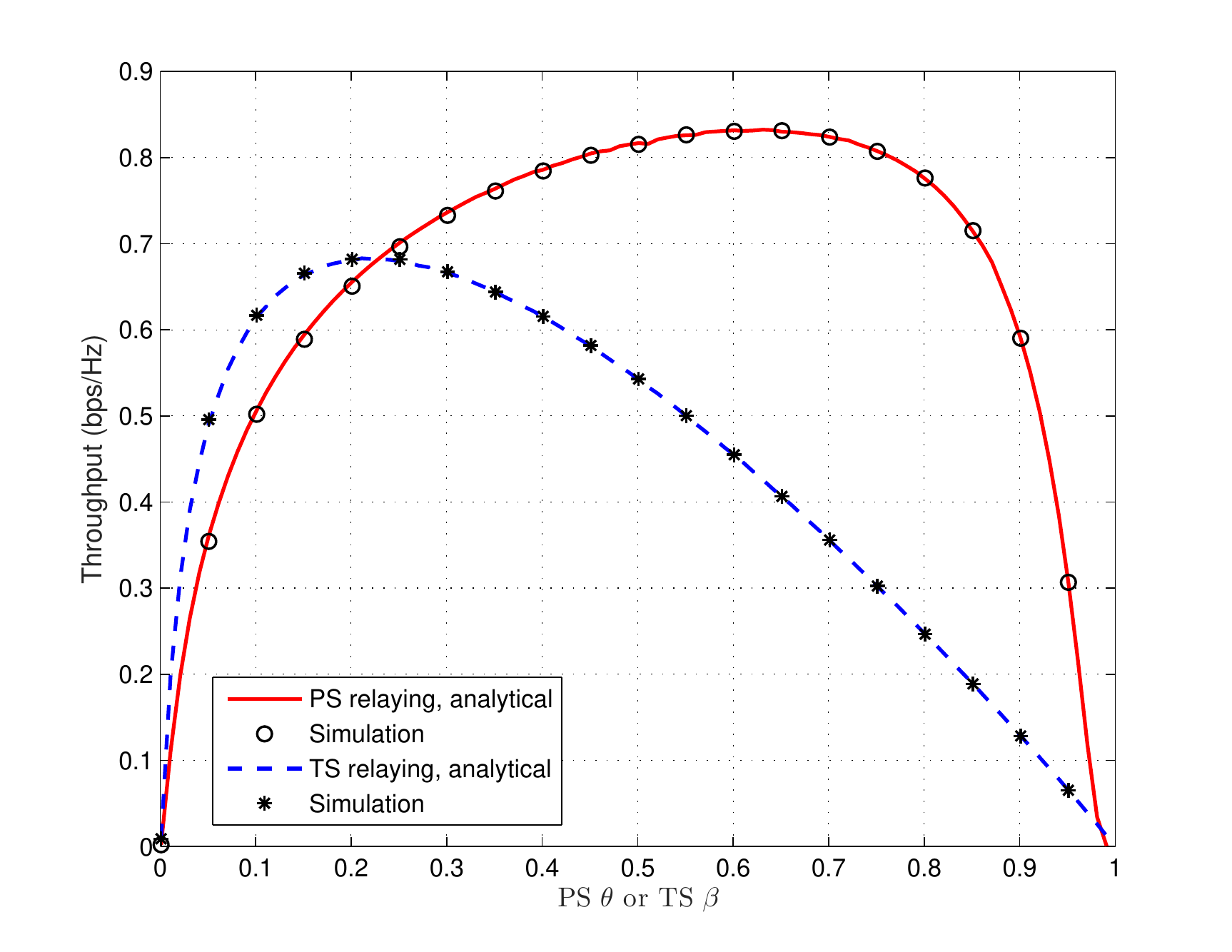}
   \caption{Throughput, $\tau$, at $D$ with respect to $\beta$ for the TS protocol and $\theta$ for the PS protocol for SNR = 20 dB. $\eta = 1$, $P_s=1$, and $d_{sr}=d_{rd}=1$.}
   \label{f6} 
   \vspace*{-1cm}
\end{figure}

 In Figs. \ref{f5} and \ref{f6}, we analyse the impact of $\beta$ and $\theta$ on the outage probability, $P_{out}$, and the achievable throughput, $\tau$, of the system, when the SNR is fixed to 20 dB. As it can be seen from Figs. \ref{f5} and \ref{f6}, there is a perfect match between the simulation and the analytical results for the entire range of the values of $\beta$ or $\theta$, indicating the high accuracy of the PA method applied to approximate the MGF of the receive SNR. It can be also observed that unlike the TS protocol, the PS protocol has an optimal value for $\theta$ that minimises the outage probability. While Fig. \ref{f6} demonstrates that there exists an optimal value for $\beta$ and $\theta$ maximising the throughput. This is because, for the PS protocol, as $\theta$ increases from 0 to an optimal value, ($\theta=0.631$), more power is exploited for EH and consequently, the relay node transmits with a higher power. Therefore, the outage probability is reduced and, yielding higher values of throughput to be observed at the destination. However, as $\theta$ increases from its optimal value, more power is consumed on EH and less power is left for source to relay information processing. As a result of weak signal strength received at the relay, the relay amplifies the noisy signal and forwards it to the destination, leading to a higher outage probability and smaller throughput to occur at the destination node. On the contrary, as $\beta$ increases from 0 to 1 for the TS protocol, more time is spent on EH and, thus, higher power is available at the relay and less outage probability is observed at the destination node as a result of a higher receive SNR. However, the performance of the system in terms of the throughput has a similar trend to that marked for the PS protocol. In particular, as $\beta$ increases from 0 to an optimal value ($\beta= 0.221$), more time is allocated for EH causing higher power to be available for information relaying and, thus, higher throughput is observed at the destination node. On the other hand, as $\beta$ increases above its optimal value, less time is available for information transmission due to a smaller value of $(1-\beta )/2$, and as a consequence, the throughput observed at the destination node decreases. 
\begin{figure}[t!]
\centering
   \includegraphics[width=3.3in]{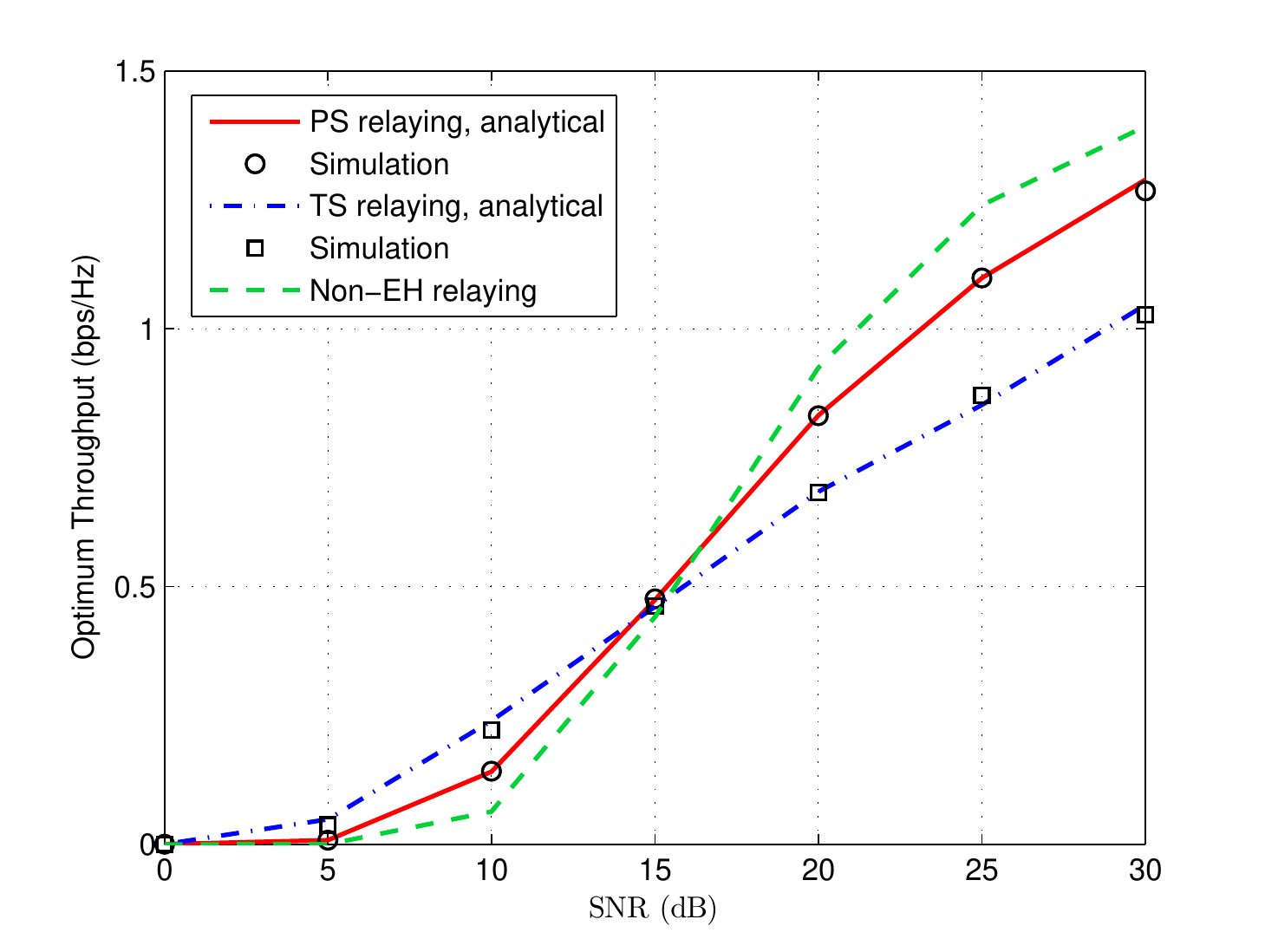}
   \caption{Optimal throughput, $\tau$, for the TS and PS protocols with respect to SNR. $\eta = 1$, $P_s=1$, $R_T=3$ bps/Hz, and $d_{sr}=d_{rd}=1$.}
   \label{f7} 
   \vspace*{-1cm}
\end{figure} 

\par \textit{Remark 1: In order to carry out a fair performance comparison between the TS and the PS protocols, it is desirable to find the values of $\beta$ and $\theta$ that maximises/minimises the value of the performance metric, i.e., throughput or ASER. Due to the PA approximation of the MGF of the receive SNR, the task of finding closed-form expressions for the optimal values of $\beta$ and $\theta$ seems intractable. Nevertheless, the optimisation can be done offline by numerically computing the optimal values of $\beta$ and $\theta$ that maximises/minimises the value of the throughput/ASER, for certain given system parameters, including, EH efficiency $\eta$, source transmission rate $R_T$, source power $P_s$, $S \to R$ distance $d_{sr}$, $R\to D$ distance, $d_{rd}$, and SNR value.}  
\par In order to further observe the effect of the PS coefficient, $\theta$, and EH time ratio, $\beta$, on the two relaying protocols, in Fig. \ref{f7}, we examine the optimal throughput, $\tau$, for different values of SNR such that the performance gain of one relaying scheme over the other can be quantified at any target throughput. Furthermore, the conventional grid-powered non-EH relaying system is plot as a benchmark\footnote{The powers applied to a conventional relay non-EH system are $P_s/2$ at $S$ and $P_s/2$ at $R$. This ensures a fair comparison with the EH system and that the total energy supply of both networks remains the same and equal to that of the noncooperative non-EH scheme.}. Fig. \ref{f7} shows that at lower SNR values (from 0 to 15 dB) both TS and PS schemes outperform the the conventional non-EH AF relaying protocol, assuming that \textit{noncoherent} modulation is applied. This stems from the fact that the SWIPT relaying system is able to boost its overall performance over the non-EH system by allowing the relay to harness extra energy from the source RF signals while accounting for the energy-rate tradeoff. However, it can be also noticed that the non-EH relaying scheme outperforms the two SWIPT relaying protocols as the SNR exceeds 15 dB. This is due to the effect of cascaded fading resulting from instantaneous EH as shown in the numerator of the receive SNR expression in \eqref{uniSNR}. Comparing the TS and the PS protocols, Fig. \ref{f7} illustrates that the TS protocol is superior to the PS protocol in achieving higher values of the throughput in the low SNR regime. This result is consistent with the one presented in \cite{Nasir2013} for \textit{coherent} modulation in SWIPT relaying systems, where the same instantaneous EH assumption is considered for a single relay. 
\par
\begin{figure}[!t]
\centering
   \includegraphics[width=3.3in]{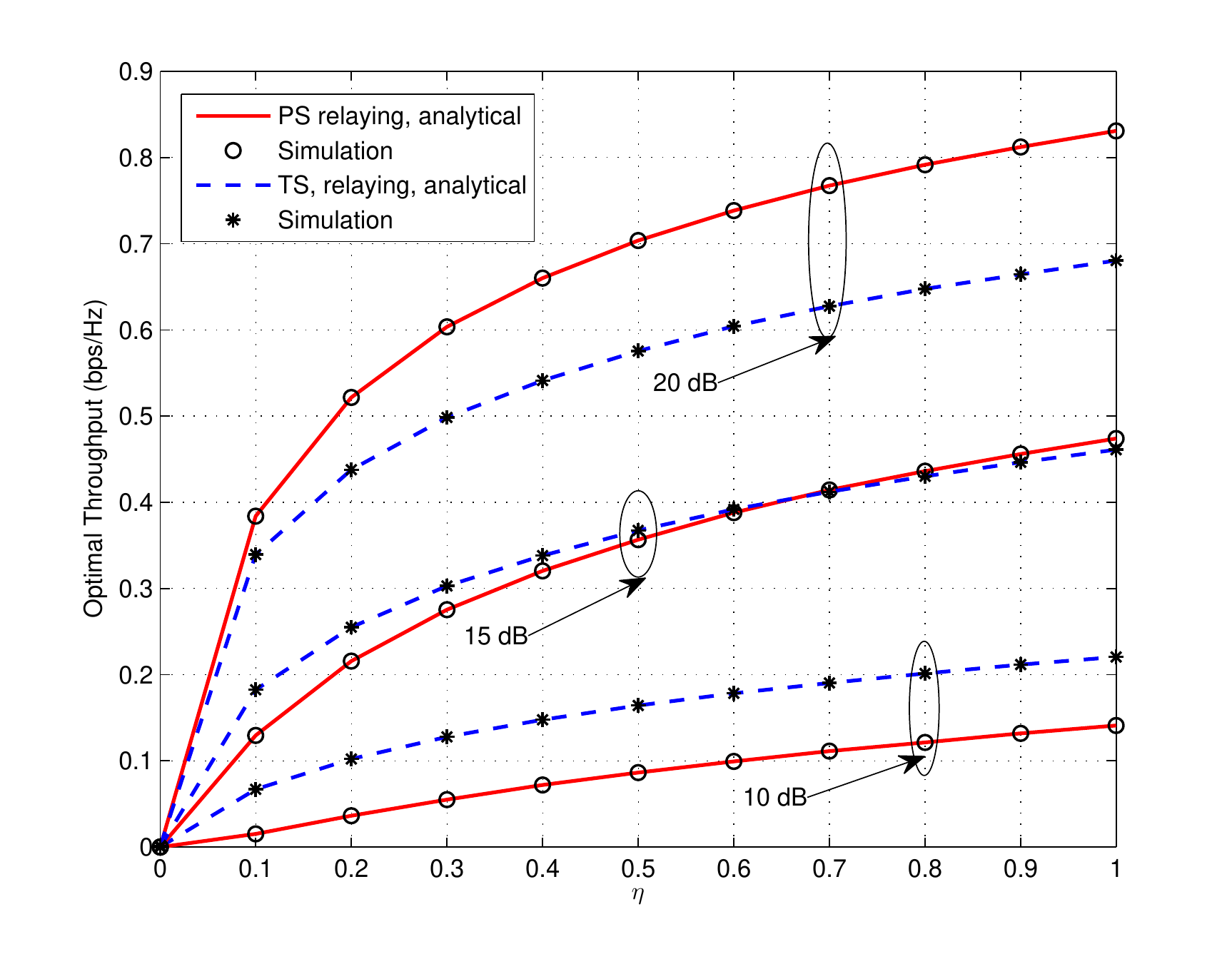}
   \caption{Optimal throughput, $\tau$, with respect to the EH efficiency, $\eta$, for the TS and PS protocols, respectively, at SNR = 10, 15, and 20 dB. $P_s=1$, $R_T=3$ bps/Hz, and $d_{sr}=d_{rd}=1$.}
   \label{f9} 
   \vspace*{-1cm}
\end{figure}

 Fig. \ref{f9} presents the optimal throughput, $\tau$, for both the TS and the PS protocols with respect to the EH efficiency, $\eta$, for three different values of the SNR, i.e., SNR = 10, 15, and 20 dB. It can be observed that at a high SNR value (SNR = 20 dB), the PS protocol outperforms the TS protocol for the entire range, while it is vice versa at a low SNR value (SNR = 10 dB). However, at SNR = 15 dB, which is the intersection point shown in Fig. \ref{f7}, the TS protocol outperforms the PS protocol when $\eta \leq 0.6$ and as $\eta$ increases beyond 0.6, the throughput performance gap between the two protocols becomes almost insignificant.    
\par
\begin{figure*}[!t]
\centering
   \begin{subfigure}[b]{0.5\textwidth}
   \includegraphics[width=\columnwidth]{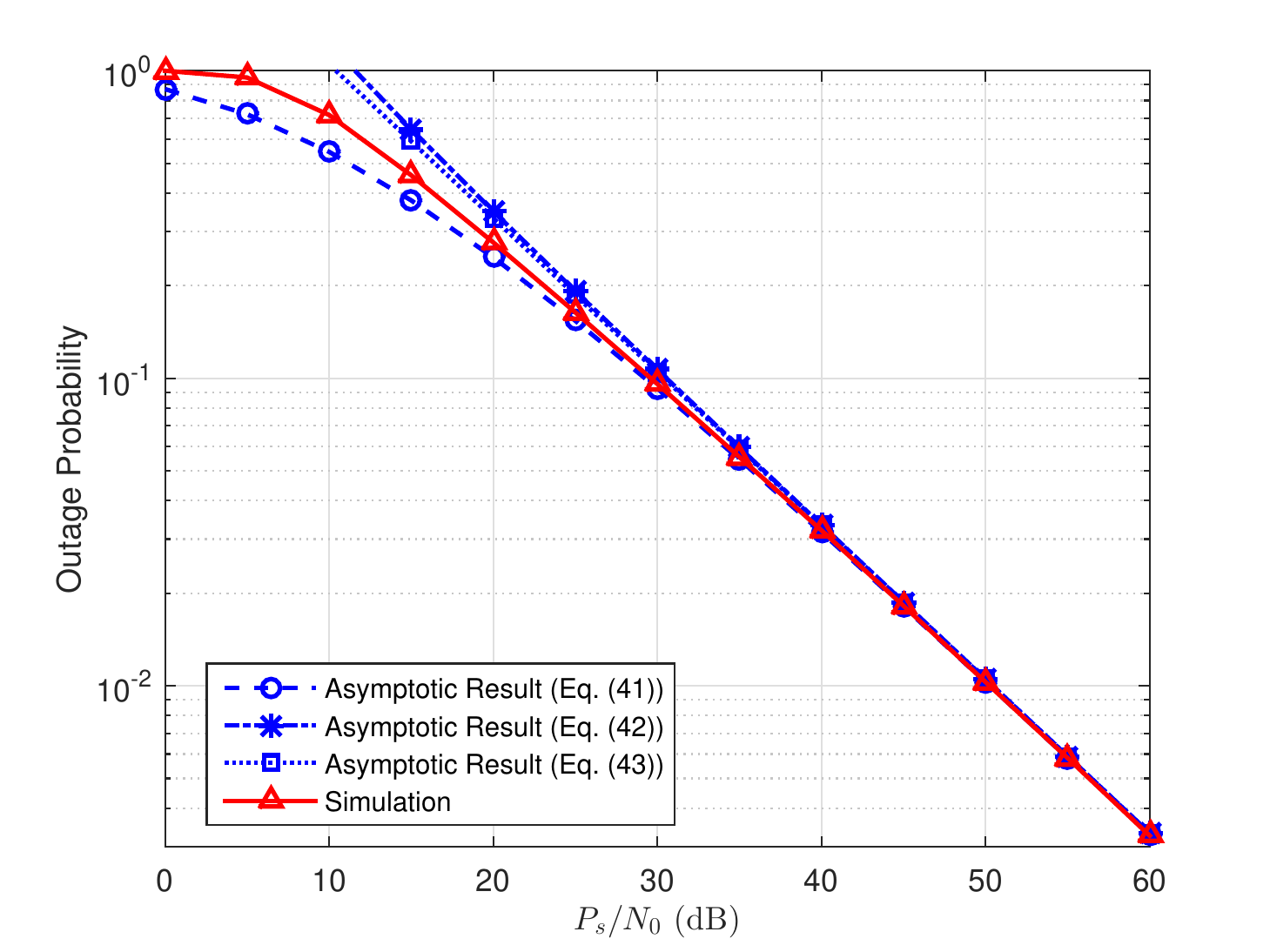}
   \caption{TS Protocol}
   \label{f3_a} 
   \vspace*{-0.5cm} 
\end{subfigure}\hfill
\begin{subfigure}[b]{0.5\textwidth}
 \includegraphics[width=\columnwidth]{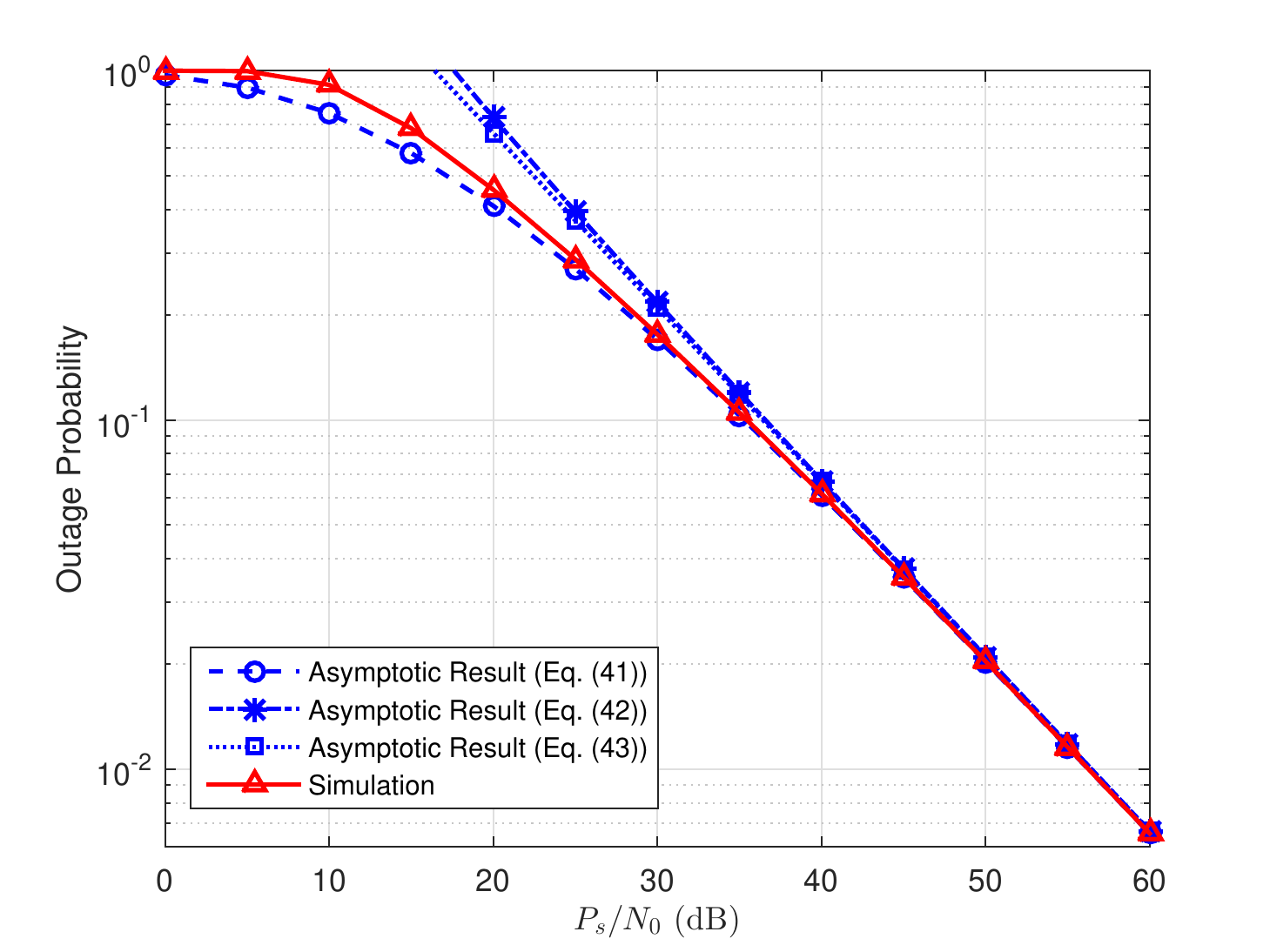}
     \caption{PS Protocol}
   \label{f3_b} 
   \vspace*{-0.5cm} 
 \end{subfigure}
 \caption{Outage probability asymptotic results with respect to SNR, $\eta = 1$, $P_s=1$, $R_T=3$ bps/Hz, and $d_{sr}=d_{rd}=1$.}
\vspace*{-1cm} 
 \label{f3} 
\end{figure*}

 In Fig. \ref{f3}, we plot the asymptotic outage probability results derived in \eqref{OPhighSR}, \eqref{OPhighSRhighRD}, and \eqref{dominantOP} against the SNR, for both the TS ($\beta=0.5$) and PS ($\theta=0.5$) relaying protocols. It is observed that all three asymptotic results perfectly match the simulation results in the high SNR regime, proving their high accuracy and significant simplicity in computing the outage probability in this region. 

\begin{figure}[t!]
\centering
   \includegraphics[width=3.3in]{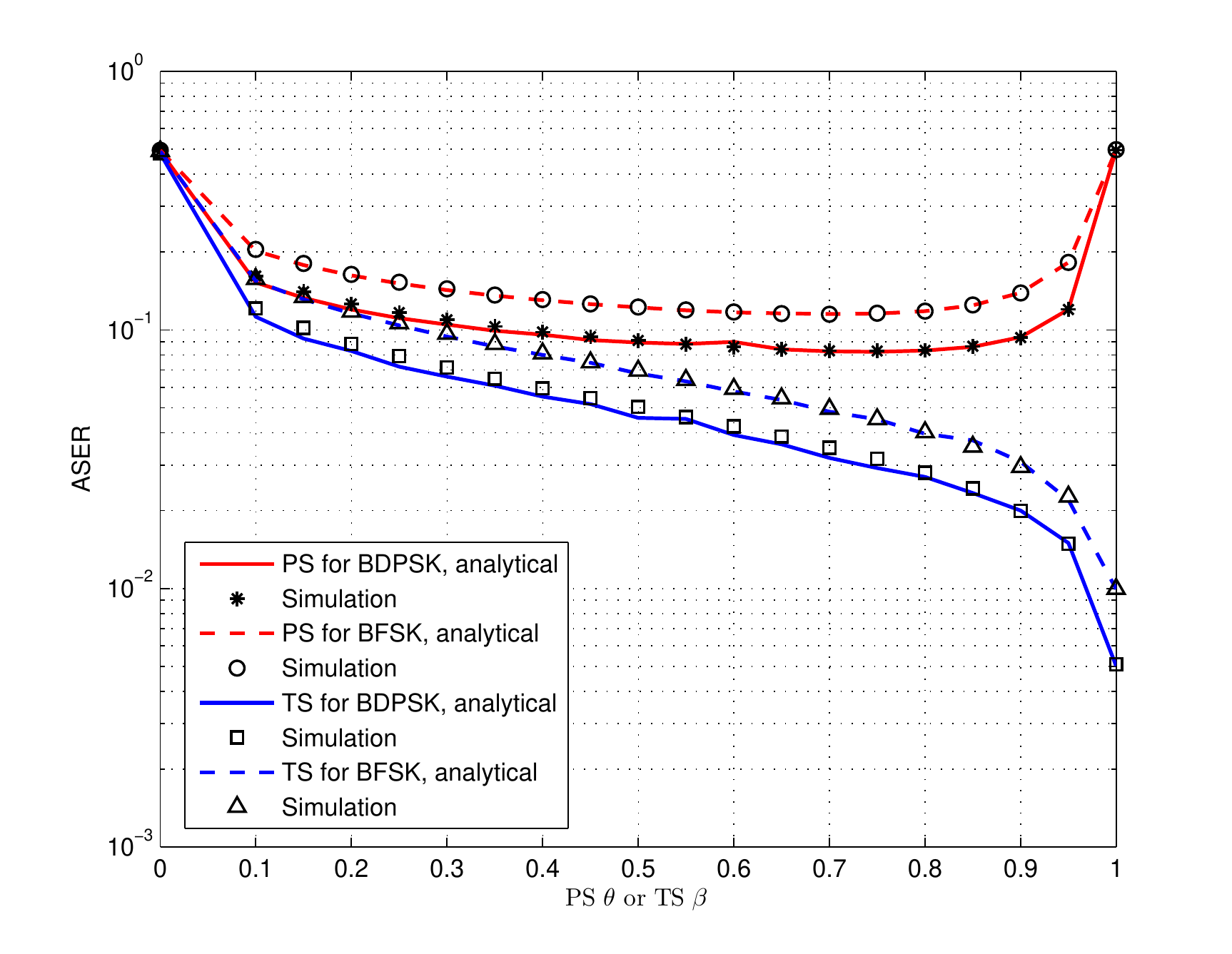}
   \caption{ASER with respect to $\beta$ or $\theta$ for TS or PS protocols, respectively, employing BDPSK and BFSK at SNR = 20 dB. $\eta = 1$, $P_s=1$, $R_T=3$ bps/Hz, and $d_{sr}=d_{rd}=1$.}
   \label{f10} 
   \vspace*{-1cm}
\end{figure} 

\par  Fig. \ref{f10} presents the ASER performance of BDPSK and binary FSK (BFSK) vs. $\beta$ and $\theta$ for the TS and the PS protocols, respectively, while fixing the SNR at 20 dB. The analytical results are obtained by computing \eqref{ber1 DPSK} and \eqref{ber1 fsk} for BDPSK and BFSK, respectively ($M=2$). Several important performance insights can be extracted from Fig. \ref{f10}. First, it is observed that the TS protocol outperforms the PS protocol across the entire range of $\beta$ or $\theta$ values. Second, it can be noticed that for binary signaling, the performance of the \textit{noncoherent} BFSK scheme is inferior to the BDPSK scheme for both TS and PS protocols. Third, Fig. \ref{f10} demonstrates that there exists a unique optimum value for the PS protocol at which the ASER value is minimised, while such an optimum value does not exist for the TS protocol.

\begin{figure*}[!t]
\centering
   \begin{subfigure}[b]{0.5\textwidth}
   \includegraphics[width=\columnwidth]{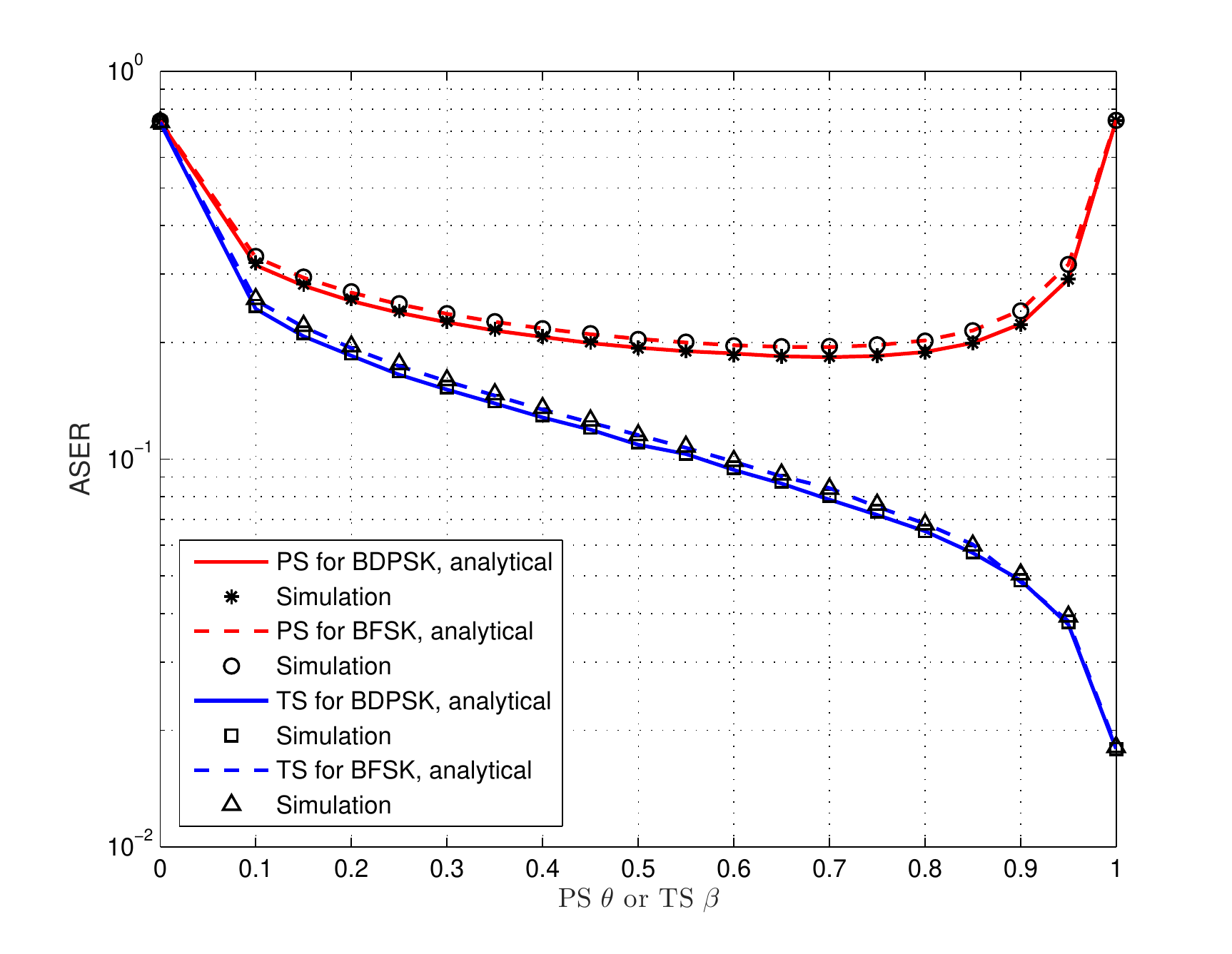}
   \caption{}
   \label{f13_a} 
   \vspace*{-0.5cm} 
\end{subfigure}\hfill
\begin{subfigure}[b]{0.5\textwidth}
 \includegraphics[width=\columnwidth]{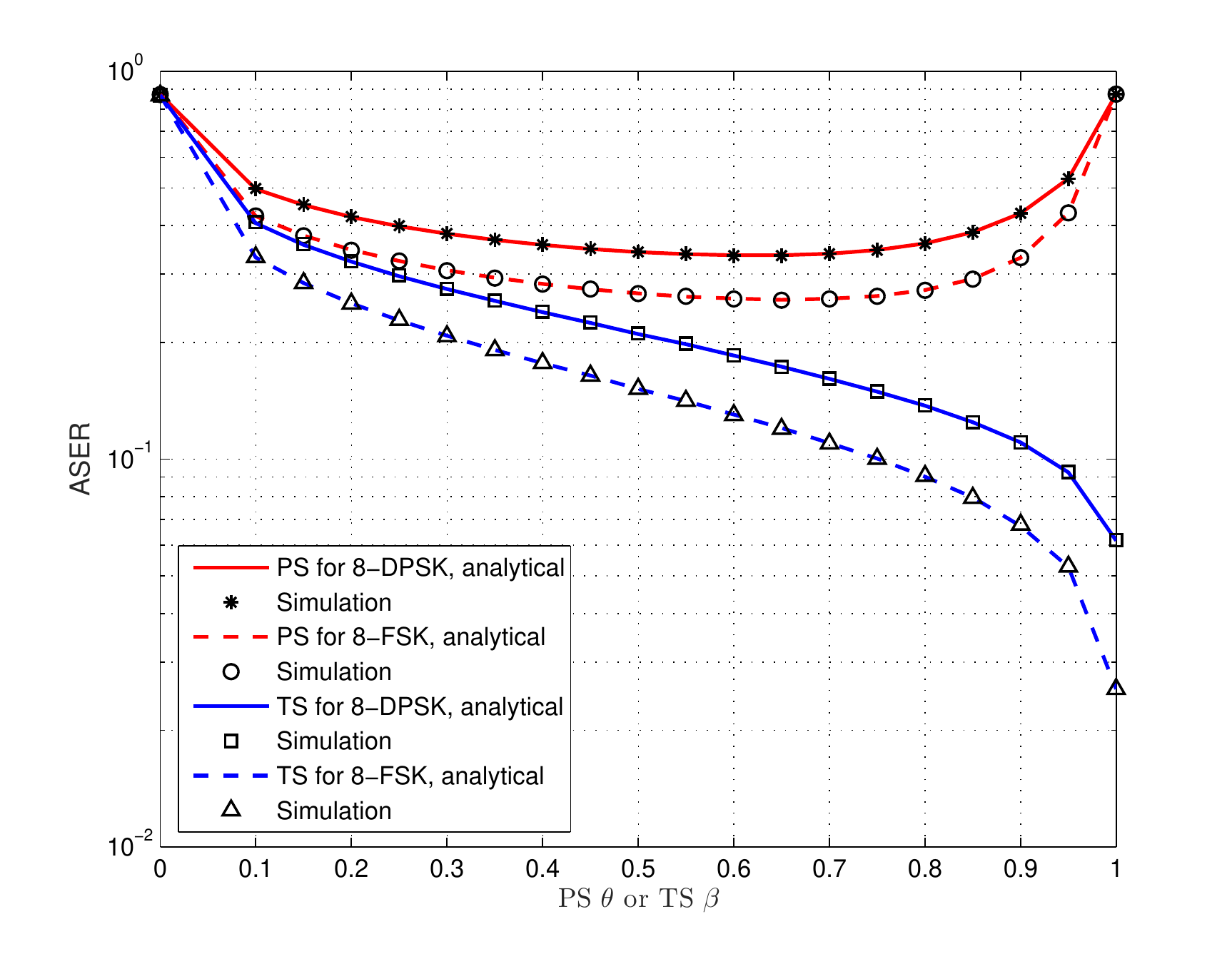}
     \caption{}
   \label{f13_b} 
   \vspace*{-0.5cm} 
 \end{subfigure}
 \caption{(a) ASER with respect to $\beta$ or $\theta$ for TS or PS protocols, respectively, employing 4-DPSK and 4-FSK. (b) ASER with respect to $\beta$ or $\theta$ for TS or PS protocols,  respectively, employing 8-DPSK and 8-FSK. For both (a) and (b) SNR = 20 dB, $\eta = 1$, $P_s=1$, $R_T=3$ bps/Hz, and $d_{sr}=d_{rd}=1$.}
\vspace*{-1cm} 
\end{figure*} 
\par
\begin{figure*}[!t]
\centering
   \begin{subfigure}[b]{0.5\textwidth}
   \includegraphics[width=\columnwidth]{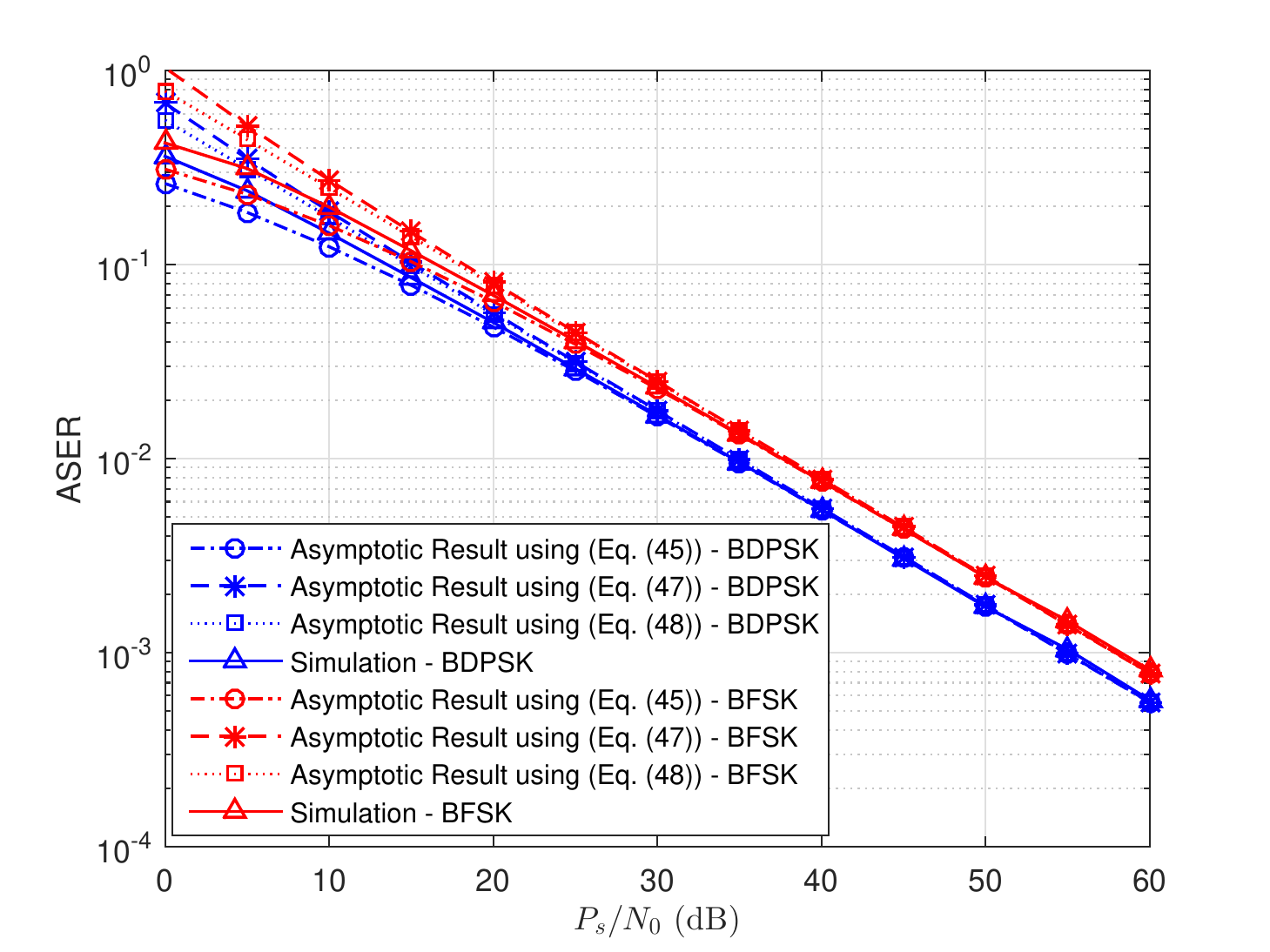}
   \caption{TS Protocol}
   \label{f4_a} 
   \vspace*{-0.5cm} 
\end{subfigure}\hfill
\begin{subfigure}[b]{0.5\textwidth}
 \includegraphics[width=\columnwidth]{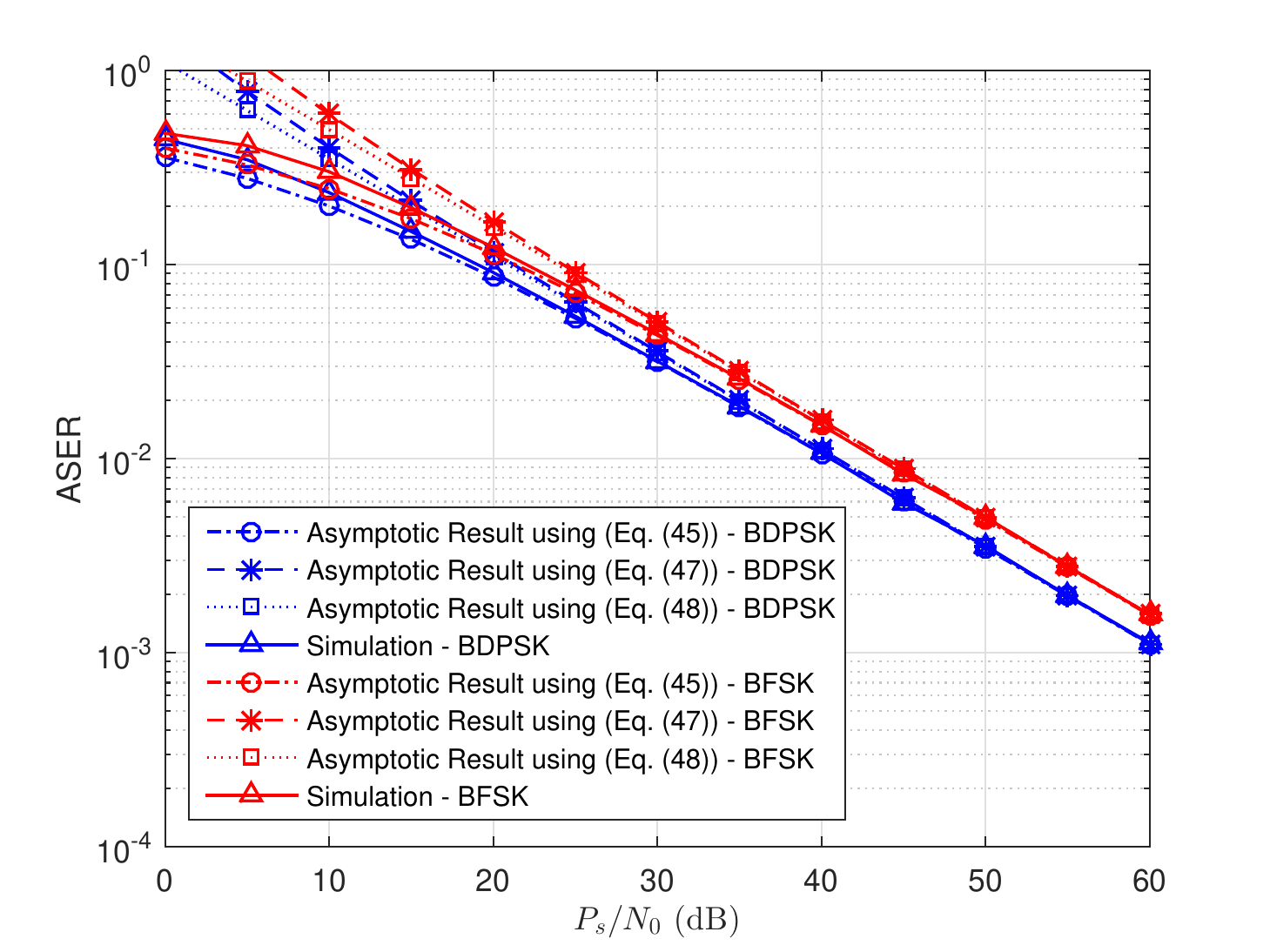}
     \caption{PS Protocol}
   \label{f4_b} 
   \vspace*{-0.5cm} 
 \end{subfigure}
 \caption{ASER asymptotic results with respect to SNR, $\eta = 1$, $P_s=1$, $R_T=3$ bps/Hz, and $d_{sr}=d_{rd}=1$.}
\vspace*{-1cm} 
 \label{f4}
\end{figure*}

 To examine the impact of higher constellation sizes $M>2$ on the ASER performance of the adopted \textit{noncoherent} $M$-DPSK and $M$-FSK signalings, we depict in Figs. \ref{f13_a} and \ref{f13_b}, the ASER performance of the TS and PS protocols for 4-DPSK and 4-FSK, and 8-DPSK and 8-FSK, respectively, when the SNR is fixed to 20 dB. For the case of $M=4$, Fig. \ref{f13_a} illustrates that 4-DPSK slightly outperforms 4-FSK, irrespective of the SWIPT relaying protocol, i.e., TS or PS. As $M$ is increased from 4 to 8, it can be observed from Fig. \ref{f13_b} that 8-FSK significantly outperforms 8-DPSK, irrespective of the SWIPT relaying protocol. The preceding results suggest that when $M \geq 8$, $M$-FSK becomes more energy efficient in terms of energy consumption than $M$-DPSK. This is also reported in \cite{Liu}.  
Finally, in Fig. \ref{f4} we present the asymptotic ASER results derived using the MGF asymptotic expressions in \eqref{SERhighSR}, \eqref{SERhighSRhighRD}, and \eqref{MGFF} with respect to the SNR for the TS ($\beta = 0.5$) and PS ($\theta = 0.5$) relaying protocols considering BFSK and BDPSK. It is shown that all asymptotic results fully coincide with the simulation results in the high SNR regime. This verifies the accuracy and efficiency of these results in quantifying the ASER performance of the system in this region.   

\section{Conclusions}\label{sec:conc}
In this paper, we presented unif\mbox{}ied analytical expressions for the moments and the MGF of the end-to-end SNR of a \textit{noncoherent} SWIPT dual-hop relay system, adopting the TS or the PS as the receiver architecture at the relay node. Capitalising on these expressions, we proposed new unified formulas for various performance metrics, including the outage probability, achievable throughput, and ASER of two \textit{noncoherent} modulation schemes. Further, we derived and presented novel asymptotic expressions for the outage probability and ASER. Additionally, we analytically demonstrated that the diversity order of the considered system is less than 1. The proposed mathematical framework demonstrates the unif\mbox{}ication of the PS and TS relaying protocols with $M$-FSK and $M$-DPSK modulation schemes into a single expression allowing one to utilise this unif\mbox{}ied expression and derive the required expression for a variety of performance metrics. 
Furthermore, the offered unif\mbox{}ied asymptotic results are applicable beyond the scope of this paper and opens the door for simple further analysis of a wide array of other scenarios and performance metrics, such as ergodic capacity. The analytical model was corroborated with computer simulations. Our results demonstrated that there is a unique value for the PS ratio that minimises the outage probability of the system, while such a value does not exist for the TS protocol. We also showed that, considering the system throughput, the TS relaying scheme is superior to the PS relaying scheme at lower SNR values. 

\appendices
 \section{Proof of Theorem I}
  \label{Appendix A}
  \par The $n^\text{th}$-order moment of the receive SNR in \eqref{uniSNR} can be calculated as 
 \begin{subequations}
\begin{align}
\mu_n&=\int_0^\infty\int_0^\infty \left(\frac{\hat{a}\hat{b}  \gamma_{sr}^2 \gamma_{rd}}{\hat{b} \gamma_{sr}\gamma_{rd} +C}\right)^n f_{\gamma_{sr}}(\gamma_{sr})f_{\gamma_{rd}}(\gamma_{rd})\text{ }d\gamma_{sr}\text{}d\gamma_{rd} \label{moment1} \\
&=\frac{1}{\lambda_{{sr}}\lambda_{{rd}}}\int_0^\infty\underbrace{\int_0^\infty \left(\frac{\hat{a}\hat{b}  \gamma_{sr}^2 \gamma_{rd}}{\hat{b} \gamma_{sr}\gamma_{rd} +C}\right)^n \text{exp}\left(-\frac{\gamma_{sr}}{\lambda_{{sr}}}\right)d\gamma_{sr}}_{I_1} \text{exp}\left(-\frac{\gamma_{rd}}{\lambda_{{rd}}}\right)\text{}d\gamma_{rd} \label{moment2} \\
&= \frac{\hat{a}^n}{\lambda_{{sr}}\lambda_{{rd}}}\int_0^\infty\underbrace{\int_0^\infty \gamma_{sr}^n\left(1+\frac{C}{\hat{b}\gamma_{sr}\gamma_{rd}}\right)^{-n} \text{exp}\left(-\frac{\gamma_{sr}}{\lambda_{{sr}}}\right)d\gamma_{sr}}_{I_1}\text{exp}\left(-\frac{\gamma_{rd}}{\lambda_{{rd}}}\right)\text{}d\gamma_{rd} \label{moment3} \\
&= \frac{\hat{a}^n}{\lambda_{{sr}}\lambda_{{rd}}\Gamma(n)}\int_0^\infty\underbrace{\int_0^\infty \gamma_{sr}^n G_{1, 1}^{1, 1}\left[\frac{\hat{b}\gamma_{sr}\gamma_{rd}}{C} \  \big\vert \  {1 \atop n} \right] G_{0, 1}^{1, 0}\left[\frac{\gamma_{sr}}{\lambda_{{sr}}} \  \Big\vert \  {- \atop 0} \right]d\gamma_{sr}}_{I_1}\text{exp}\left(-\frac{\gamma_{rd}}{\lambda_{{rd}}}\right)\text{}d\gamma_{rd} \label{moment4} \\
&=\frac{\hat{a}^n}{\lambda_{{sr}}\lambda_{rd}\Gamma(n)}\left(\frac{\hat{b}}{C}\right)^{-n-1}\int_0^\infty \gamma_{rd}^{-n-1}G^{1, 2}_{2, 1}\left[\frac{\hat{b}\lambda_{{sr}}\gamma_{rd}}{C} \  \big\vert \  {1, n+2 \atop 2n+1}\right]\text{exp}\left(-\frac{\gamma_{rd}}{\lambda_{{rd}}}\right)\text{}d\gamma_{rd},\label{moment5} 
\end{align}
\end{subequations}
where $C = \hat{a}\sigma^2_{sr}+ \Psi$ and $\Gamma(.)$ is the Gamma function \cite[Eq. (8.310.1)]{Rizhik}. Hereby, \eqref{moment2} follows because $f_{\gamma_{sr}}(\gamma_{sr})$ and $f_{\gamma_{rd}}(\gamma_{rd})$ are the PDFs of the exponential random variables $\gamma_{sr}$ and  $\gamma_{rd}$, with means $\lambda_{sr}$ and $\lambda_{rd}$, respectively. Furthermore, \eqref{moment4} is obtained by expressing the integrands of $I_1$ in \eqref{moment3} in terms of their Meijer G-function representations. Specifically, the second integrand of $I_1$ can be expressed as $\frac{1}{\Gamma(n)}G_{1, 1}^{1, 1}\left[\frac{\hat{b}\gamma_{sr}\gamma_{rd}}{C} \  \vert \  {1 \atop n} \right]$, where the equality in \cite[Eq. (8.4.2.5)]{Prudnikov} is used followed by applying the transformation \cite[Eq. (8.2.2.14)]{Prudnikov}. Also, the third integrand is rewritten by making use of the equality $e^{-\frac{\gamma_{sr}}{\lambda_{{sr}}}}=G_{0, 1}^{1, 0}\left[\frac{\gamma_{sr}}{\lambda_{{sr}}} \  \vert \  {- \atop 0} \right]$\cite[Eq. (8.4.3.1)]{Prudnikov}. Then, by exploiting the integral identity \cite[Eq. (2.24.1.1)]{Prudnikov} followed by performing some algebraic manipulations, $I_1$ can be derived in a closed-form as in \eqref{moment5}. Finally, by expressing $e^{-\frac{\gamma_{rd}}{\lambda_{{rd}}}} = G_{0, 1}^{1, 0}\left[\frac{\gamma_{rd}}{\lambda_{{rd}}} \  \vert \  {- \atop 0} \right]$ and using again the aid of \cite[Eq. (2.24.1.1)]{Prudnikov}, the desired result in \eqref{moment final} is reached.

\section{Proof of Proposition 1}
  \label{Appendix B}
The outage probability of the approximated $\gamma_{eq}$ can be expressed as
\begin{subequations}
\begin{align}
P_{{out1,1}}^\infty&\cong  \text{Pr}\left(\hat{b}\gamma_{sr}^2\gamma_{rd} \leq \gamma_{th}\right) \\
&\cong \text{Pr}\left(\gamma_{rd} \leq \frac{\gamma_{th}}{\hat{b}\gamma_{sr}^2}\right) \\
&\cong\int_0^\infty f_{\gamma_{sr}}(\gamma_{sr})\left(1-\text{exp}\left(-\frac{\gamma_{th}}{\hat{b}\lambda_{rd}\gamma_{sr}^2}\right)\right)d\gamma_{sr}  \label{eq3} \\
&\cong 1-\frac{1}{\lambda_{sr}}\int_0^\infty \text{exp}\left(-\frac{\gamma_{sr}}{\lambda_{sr}}\right)\text{exp}\left(-\frac{\gamma_{th}}{\hat{b}\lambda_{rd}\gamma^2_{sr}}\right)d\gamma_{sr}  \label{eq4} \\
&\cong 1-\frac{1}{\lambda_{sr}}\int_0^\infty \text{exp}\left(-\frac{\gamma_{sr}}{\lambda_{sr}}\right) G^{0, 1}_{1, 0}\left[\frac{\hat{b}\lambda_{{rd}}\gamma_{sr}^2}{\gamma_{th}} \  \big\vert \  {1 \atop -}\right] d\gamma_{sr}, \label{eq5}
\end{align}
\end{subequations} where \eqref{eq4} follows from \eqref{eq3} because $f_{\gamma_{sr}}(\gamma_{sr})$ is the PDF of the exponential random variable $\gamma_{sr}$. Moreover, \eqref{eq5} is obtained by using the identity \cite[Eq. (8.4.3.1)]{Prudnikov} to express the second integrand of \eqref{eq4} in terms of its Meijer G-function representation whose arguments are then inverted by applying the transformation \cite[Eq. (8.2.2.14)]{Prudnikov}. Finally, by making use of the identity in \cite[Eq. (7.813.2)]{Rizhik}, we derive the desired result in \eqref{OPhighSR}.

\renewcommand{\baselinestretch}{1.42}
\bibliographystyle{IEEEtran}
\bstctlcite{BSTcontrol}
\bibliography{Reflist}

\begin{thebibliography}{10}
\providecommand{\url}[1]{#1}
\csname url@samestyle\endcsname
\providecommand{\newblock}{\relax}
\providecommand{\bibinfo}[2]{#2}
\providecommand{\BIBentrySTDinterwordspacing}{\spaceskip=0pt\relax}
\providecommand{\BIBentryALTinterwordstretchfactor}{4}
\providecommand{\BIBentryALTinterwordspacing}{\spaceskip=\fontdimen2\font plus
\BIBentryALTinterwordstretchfactor\fontdimen3\font minus
  \fontdimen4\font\relax}
\providecommand{\BIBforeignlanguage}[2]{{%
\expandafter\ifx\csname l@#1\endcsname\relax
\typeout{** WARNING: IEEEtran.bst: No hyphenation pattern has been}%
\typeout{** loaded for the language `#1'. Using the pattern for}%
\typeout{** the default language instead.}%
\else
\language=\csname l@#1\endcsname
\fi
#2}}
\providecommand{\BIBdecl}{\relax}
\BIBdecl

\bibitem{Mohjazi3}
\BIBentryALTinterwordspacing
L.~Mohjazi, S.~Muhaidat, M.~Dianati, and M.~Al-Qutayri, ``Outage probability
  and throughput of {SWIPT} relay networks with differential modulation,''
  \emph{accepted to appear in Proc. IEEE VTC Fall}, 2017. [Online]. Available:
  \url{https://arxiv.org/abs/1702.03692}
\BIBentrySTDinterwordspacing

\bibitem{Bi}
S.~Bi, C.~K. Ho, and R.~Zhang, ``Wireless powered communication: opportunities
  and challenges,'' \emph{IEEE Commun. Magazine}, vol.~53, no.~4, pp. 117--125,
  2015.

\bibitem{Mohjazi2}
L.~Mohjazi, M.~Dianati, G.~Karagiannidis, S.~Muhaidat, and M.~Al-Qutayri,
  ``{RF}-powered cognitive radio networks: technical challenges and
  limitations,'' \emph{IEEE Commun. Mag.}, vol.~53, no.~4, pp. 94--100, 2015.

\bibitem{Lmohjazi}
L.~Mohjazi, I.~Ahmed, S.~Muhaidat, M.~Dianati, and M.~Al-Qutayri, ``Downlink
  beamforming for {SWIPT} multi-user {MISO} underlay cognitive radio
  networks,'' \emph{IEEE Commun. Lett.}, vol.~21, no.~2, pp. 434--437, 2017.

\bibitem{Varshney2008}
L.~R. Varshney, ``Transporting information and energy simultaneously,'' in
  \emph{IEEE Int. Symp. Inf. Theory ({ISIT}'08)}, Toronto, July 2008, pp.
  1612--1616.

\bibitem{Grover}
P.~Grover and A.~Sahai, ``Shannon meets tesla: Wireless information and power
  transfer,'' in \emph{IEEE Int. Symp. Int. Theory ({ISIT}'10)}, June 2010, pp.
  2363--2367.

\bibitem{Nasir2013}
A.~Nasir, X.~Zhou, S.~Durrani, and R.~Kennedy, ``Relaying protocols for
  wireless energy harvesting and information processing,'' \emph{IEEE Trans.
  Wireless Commun.}, vol.~12, no.~7, pp. 3622--3636, Jul. 2013.

\bibitem{Rabie}
K.~M. Rabie, A.~Salem, E.~Alsusa, and M.~S. Alouini, ``Energy-harvesting in
  cooperative {AF} relaying networks over log-normal fading channels,'' in
  \emph{IEEE Int. Conf. on Commun. ({ICC}'16)}, 2016, pp. 1--7.

\bibitem{Zhiguo1}
Z.~Ding, I.~Krikidis, B.~Sharif, and H.~Poor, ``Wireless information and power
  transfer in cooperative networks with spatially random relays,'' \emph{IEEE
  Trans. Wireless Commun.}, vol.~13, no.~8, pp. 4440--4453, 2014.

\bibitem{Krikidis}
I.~Krikidis, ``Simultaneous information and energy transfer in large-scale
  networks with/without relaying,'' \emph{IEEE Trans. Commun.}, vol.~62, no.~3,
  pp. 900--912, 2014.

\bibitem{Lee}
H.~Lee, C.~Song, S.~H. Choi, and I.~Lee, ``Outage probability analysis and
  power splitter designs for {SWIPT} relaying systems with direct link,''
  \emph{IEEE Commun. Lett.}, vol.~21, no.~3, pp. 648--651, 2017.

\bibitem{Liu2010}
P.~Liu and I.~M. Kim, ``Optimum/sub-optimum detectors for multi-branch dual-hop
  amplify-and-forward cooperative diversity networks with limited {CSI},''
  \emph{IEEE Trans. Wireless Commun.}, vol.~9, no.~1, pp. 78--85, 2010.

\bibitem{Liu}
P.~Liu, S.~Gazor, I.-M. Kim, and D.~Kim, ``Noncoherent relaying in energy
  harvesting communication systems,'' \emph{IEEE Trans. Wireless Commun.},
  vol.~14, no.~12, pp. 6940--6954, 2015.

\bibitem{Gazor}
------, ``Energy harvesting noncoherent cooperative communications,''
  \emph{IEEE Trans. Wireless Commun.}, vol.~14, no.~12, pp. 6722--6737, 2015.

\bibitem{Xu}
W.~Xu, Z.~Yang, Z.~Ding, L.~Wang, and P.~Fan, ``Wireless information and power
  transfer in two-way relaying network with non-coherent differential
  modulation,'' \emph{EURASIP J. on Wireless Commun. Netw.}, no. 131, May 2015.

\bibitem{Mohjazi}
L.~Mohjazi, S.~Muhaidat, and M.~Dianati, ``Performance analysis of differential
  modulation in {SWIPT} cooperative networks,'' \emph{IEEE Signal Process.
  Lett.}, vol.~23, no.~5, pp. 620--624, 2016.

\bibitem{Lou}
Y.~Lou, Q.~Y. Yu, J.~Cheng, and H.~L. Zhao, ``Exact {BER} analysis of selection
  combining for differential {SWIPT} relaying systems,'' \emph{IEEE Signal
  Process. Lett.}, vol.~PP, no.~99, pp. 1--1, 2017.

\bibitem{Nabar}
R.~U. Nabar, H.~Bolcskei, and F.~W. Kneubuhler, ``Fading relay channels:
  Performance limits and spacetime signal design,'' \emph{EEE J. Sel. Areas
  Commun.}, vol.~22, pp. 1099--1109, Aug. 2004.

\bibitem{Zhao}
Q.~Zhao and H.~Li, ``Differential modulation for cooperative wireless
  systems,'' \emph{IEEE Trans. Signal Process.}, vol.~55, no.~5, pp.
  2273--2283, May 2007.

\bibitem{Evans}
S.~Atapattu and J.~Evans, ``Optimal power-splitting ratio for wireless energy
  harvesting in relay networks,'' in \emph{IEEE 82nd Veh. Technol. Conf.
  ({VTC}'15-{F}all)}, Sept. 2015, pp. 1--6.

\bibitem{Proakis}
J.~G. Proakis, \emph{Digital Communications}.\hskip 1em plus 0.5em minus
  0.4em\relax New York: McGraw-Hill, 4th edition, 2000.

\bibitem{Prudnikov}
A.~P. Prudnikov, Y.~A. Brychkov, and O.~I. Marichev, \emph{Integrals and
  Series}.\hskip 1em plus 0.5em minus 0.4em\relax Gordon and Breach Science
  Publishers, 1986, vol.~3.

\bibitem{Baker}
G.~A. Baker and P.~Graves-Morris, \emph{Pad{\'e} Approximants}.\hskip 1em plus
  0.5em minus 0.4em\relax Cambridge, UK.: Cambridge Univ. Press, 1996.

\bibitem{Amindavar}
H.~Amindavar and J.~A. Ritcey, ``Pad{\'e} approximations of probability density
  functions,'' \emph{IEEE Trans. Aerosp. Elect. Syst.}, vol.~30, no.~2, pp.
  416--424, 1994.

\bibitem{Simon}
M.~K. Simon and M.-S. Alouini, \emph{Digital Communications over Fading
  Channels. A Unified Approach to Performance Analysis.}\hskip 1em plus 0.5em
  minus 0.4em\relax New York, NY: John Wiley and Sons, Inc., 2000.

\bibitem{Young-Chai}
Y.-C. Ko, M.~S. Alouini, and M.~K. Simon, ``Outage probability of diversity
  systems over generalized fading channels,'' \emph{IEEE Trans. Commun.},
  vol.~48, no.~11, pp. 1783--1787, 2000.

\bibitem{Alouini}
M.~K. Simon and M.~S. Alouini, ``A unified approach to the probability of error
  for noncoherent and differentially coherent modulations over generalized
  fading channels,'' \emph{IEEE Trans. Commun.}, vol.~46, no.~12, pp.
  1625--1638, 1998.

\bibitem{Ansari}
I.~S. Ansari, F.~Yilmaz, and M.~S. Alouini, ``Performance analysis of
  free-space optical links over malaga ($\mathcal{M} $) turbulence channels
  with pointing errors,'' \emph{IEEE Trans. Wireless Commun.}, vol.~15, no.~1,
  pp. 91--102, 2016.

\bibitem{Rizhik}
I.~M. Ryzhik and I.~S. Gradshteyn, \emph{Table of Integrals, Series, and
  Products}.\hskip 1em plus 0.5em minus 0.4em\relax Academic Press, 7th
  edition, 2007.

\end{thebibliography}
\end{document}